\documentclass[sigconf,nonacm]{acmart}

\renewcommand\footnotetextcopyrightpermission[1]{}
\setcopyright{none}

\AtBeginDocument{
  }

\begin{document}

\title{ReforMe: Re-Shaping Documents with Contextual Prompting and Layout-Aware Propagation}

\author{Nabin Khanal}
\email{khanaln@purdue.edu}
\affiliation{
  \institution{Purdue University}
  \city{West Lafayette}
  \state{Indiana}
  \country{USA}
}

\author{Tongyan Wang}
\email{wang5298@purdue.edu}
\affiliation{
  \institution{Purdue University}
  \city{West Lafayette}
  \state{Indiana}
  \country{USA}
}

\author{Jui-Cheng Chiu}
\email{chiu119@purdue.edu}
\affiliation{
  \institution{Purdue University}
  \city{West Lafayette}
  \state{Indiana}
  \country{USA}
}

\author{Ningning Nicole Kong}
\affiliation{
  \institution{Purdue University}
  \city{West Lafayette}
  \state{Indiana}
  \country{USA}
}

\author{Hannah Yanhua Zong}
\email{zong6@purdue.edu}
\affiliation{
  \institution{Purdue University}
  \city{West Lafayette}
  \state{Indiana}
  \country{USA}
}

\author{Yingjie Victor Chen}
\email{victorchen@purdue.edu}
\affiliation{
  \institution{Purdue University}
  \city{West Lafayette}
  \state{Indiana}
  \country{USA}
}

\renewcommand{\shortauthors}{Khanal et al.}

\begin{abstract}
Digitizing complex documents with handwritten content, irregular tables, and heterogeneous layouts remains challenging, as traditional Optical Character Recognition (OCR) systems fail to capture writing nuances, author-specific conventions, and document structure, and recent LLM-based approaches lack mechanisms for precise, scalable correction. We present an interactive document digitization system that integrates layout-aware parsing, OCR, and LLM-based reconstruction with user-driven refinement. The system is informed by a formative study that identifies key challenges and interaction needs in real-world digitization workflows. It supports both direct edits and natural-language instructions, and introduces a layout-aware propagation mechanism that generalizes user corrections across structurally similar regions. This enables not only efficient error correction but also document re-shaping into structured, analyzable representations. We evaluate the system through a within-subjects user study (n=12) on real-world documents. Results show improved correction efficiency and reduced repetitive effort, demonstrating more effective and controllable document digitization procedure.

\end{abstract}

\keywords{document digitization, human-in-the-loop systems, large language models, layout-aware editing, correction propagation, document reformatting}

\begin{teaserfigure}
    \centering
    \includegraphics[width=\textwidth]{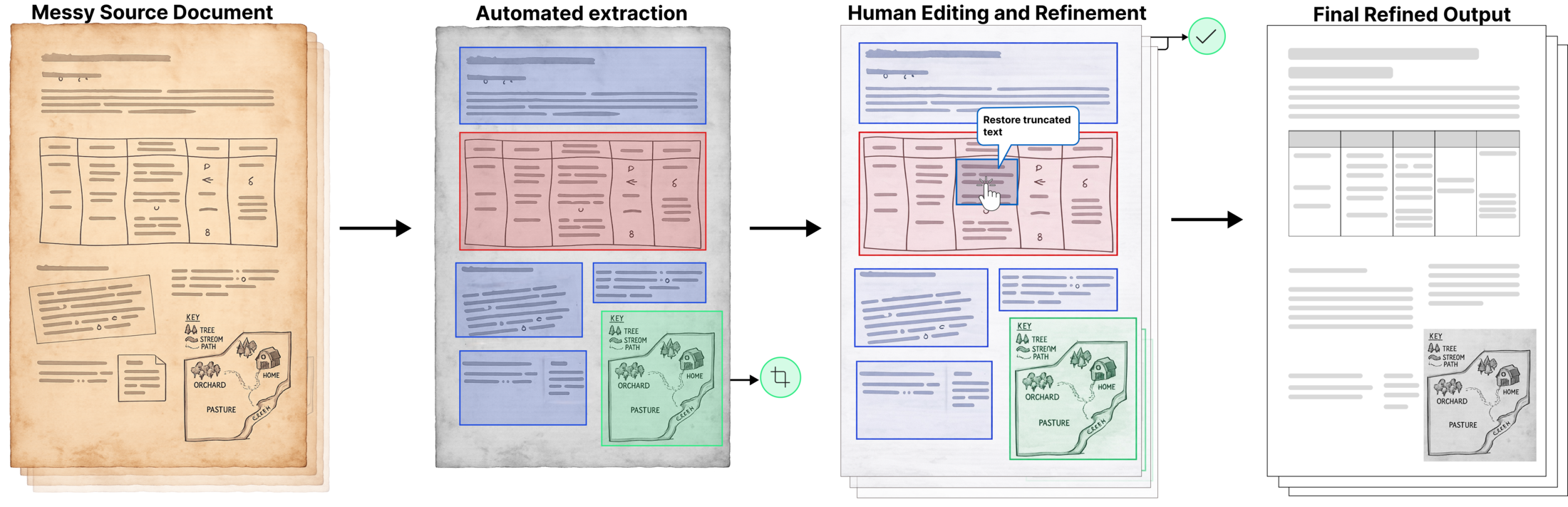}
    \caption{\textbf{From noisy scans to interactive, layout-aware refinement.}
    Starting from a messy archival source document (left), our system first performs \emph{layout parsing} to detect semantic regions such as text blocks, tables, and figures (second panel; blue boxes denote text regions, the red box denotes a table region, and the green box denotes a figure region). These detected regions are then used for \emph{region-level extraction}: text and tables are reconstructed into editable structured content, while figures can be \emph{cropped and preserved} as separate visual elements. In the third panel, users refine the extracted result through \emph{direct editing} and \emph{natural-language prompting}; for example, a user edits a selected table cell and issues the instruction \textit{``Restore truncated text''}. The green check icon indicates that the update is accepted and applied to similar sections. From this interaction, the system re-prompts the model with the original region, current extraction, and user-provided edits or instructions, then regenerates the affected section and can \emph{automatically propagate} the same inferred fix to other structurally similar regions or repeated content across pages. The final result (right) is a cleaner, more structured, and more usable digital document that preserves both reconstructed text and retained figures.}
    \Description{A four-stage teaser figure showing an archival document, layout parsing into text, table, and figure regions, human editing and prompting on a selected table region with an instruction to restore truncated text, and a final refined digital output. The green crop icon denotes localized region selection, and the green check icon denotes acceptance and application of the refinement.}
    \label{fig:teaser}
\end{teaserfigure}

\maketitle

\section{Introduction}
Traditionally, documentation has relied heavily on physical records, including handwritten field notes, hand-drawn maps, and observational logs collected across decades of fieldwork. These materials remain essential for research, planning, and decision-making, and exist in large volumes across many domains \cite{MAZURCZYK2018111}. However, accessing these documents typically requires either direct consultation or reliance on scanned images, both of which limit efficient interaction with the content, presenting significant challenges for accessibility, searchability, and integration into modern digital systems. To enable search, analysis, and integration with modern data-driven workflows, these documents must be converted into structured digital formats. However, this process remains labor-intensive and challenging, hindering large-scale digitization efforts. As a result, many records remain disconnected from modern data infrastructures, restricting large-scale analysis and limiting their broader utility in research and decision-making.

Prior research focused on digitalizing physical document has mainly employed Optical Character Recognition (OCR) system. OCR systems transform scanned images or PDFs into machine-readable text \cite{smith2007tesseract}. While traditional OCR systems perform well on clean, high-quality documents, they struggle with degraded scans, handwritten content, and complex layouts \cite{plamondon2000online}. Noise, faint text, and irregular structures often disrupt recognition and reading order, resulting in high error rates and inconsistent outputs \cite{wang2021ocrsurvey}. Moreover, these systems typically operate at the character or line level and do not preserve higher-level document structure due to their lack of context-aware understanding of document semantics, which limits their ability to reconstruct structured content accurately. As a result, spatially organized content such as tables is difficult to reconstruct accurately, requiring additional post-processing to recover meaningful structure \cite{schreiber2017deepdesrt}.

In recent years, this space has seen a shift with the emergence of deep learning and large language models (LLMs). LLMs bring capabilities such as context understanding and error correction to document digitization \cite{hu2021misspellingcorrectionpretrainedcontextual, brown2020language}. As a result, they can recover missing or degraded text and produce more coherent outputs. Despite these improvements, challenges remain. Vision-language models struggle with text-dense documents due to limited image token capacity \cite{zhang2024llavareadenhancingreadingability}. Additionally, handwritten and domain-specific documents often use abbreviations and non-standard notations, which LLMs only partially interpret \cite{li2023vision}. Documents may also contain non-text elements such as graphs and figures, requiring targeted extraction strategies \cite{katti2018chargrid}. In some cases, extracted data often requires normalization and restructuring (e.g., extracting data from a hand-drawn map) before it can be integrated into downstream systems, which often requires task-specific prompting \cite{liu2023visual}.

To address these challenges, we present a human-in-the-loop digitization pipeline that treats LLMs not as autonomous extractors, but as collaborative agents that support users in iteratively refining document reconstructions \cite{wu2021humanintheloopdocumentlayoutanalysis}. The design of our system is informed by a formative study that identifies key challenges and interaction needs in real-world digitization workflows. Our system is implemented as a web-based platform that integrates layout parsing, OCR, and LLM-based reconstruction within a unified processing pipeline. Given an uploaded PDF, the system first generates an initial structured representation, which users can refine through two complementary interaction mechanisms: direct editing of extracted content and natural-language instructions describing desired changes. These inputs are incorporated into the pipeline to apply corrections and propagate consistent edits across structurally similar regions. By grounding updates in user intent while maintaining control over scope, the system enables efficient error correction, consistency enforcement, and re-shaping of document structure. We evaluate the system through a user study to assess its effectiveness and usability in real-world scenarios.

In summary, we make the following contributions:
\begin{enumerate}
    \item An end-to-end document digitization pipeline that processes PDFs into structured representations through layout segmentation, region-level extraction, and reconstruction into HTML or Markdown.

    \item  A formative study that investigates user workflows, challenges, and interaction needs in document digitization tasks.
    \item A human-in-the-loop interaction framework that supports direct editing and instruction-based corrections, along with a propagation mechanism that generalizes user edits across structurally similar regions.
    \item A user study evaluating the usability, effectiveness, and interaction patterns of the proposed system.

\end{enumerate}

\section{Related Work}

\subsection{Automated document digitization}
Classic digitization pipelines decompose document understanding into modular stages such as page segmentation, line finding, character recognition, and downstream text processing, exemplified by widely deployed OCR engines like Tesseract \cite{smith2007tesseract}. Modern OCR has increasingly shifted toward transformer-based recognition models (e.g., TrOCR) that unify vision encoding and text generation, improving robustness on diverse fonts and noise patterns \cite{li2023trocr}. However, the core challenge for practical digitization is not only recognizing characters but also preserving the two-dimensional organization that gives documents meaning. This has driven extensive work on multimodal document AI that jointly models text content, spatial layout, and visual cues. LayoutLM-style pretraining shows that explicitly incorporating bounding boxes and image features into pretraining improves information extraction, receipt parsing, and document QA \cite{xu2021layoutlmv2,huang2022layoutlmv3}. In parallel, generative formulations have gained prominence: OCR-free models such as Donut argue that OCR cost and OCR error propagation can limit end-to-end reliability, and instead directly decode structured outputs from document images \cite{kim2022donut}. Related image-to-text pretraining (e.g., parsing screenshots into simplified HTML) further supports document conversion as a structured generation task and improves transfer across visually-situated domains \cite{lee2023pix2struct}. Recent foundation-document models extend these ideas by unifying vision, text, and layout into prompt-based sequence generation schemes and by introducing layout-aware LLM-style decoding that leverages bounding-box structure while retaining the benefits of language-model generation \cite{tang2023udop,wang2024docllm}. Despite these advances, digitization systems remain sensitive to document-domain shifts, reading-order ambiguities, and local recognition failures; large, diverse layout benchmarks such as DocLayNet highlight that even strong detectors can drop substantially when moving beyond narrow source distributions \cite{pfitzmann2022doclaynet}. These gaps motivate workflows that treat conversion as an iterative, user-steerable process, where local corrections can be applied and propagated to improve both fidelity and editability of the final structured output.

\subsection{Recovering structure in visually organized content.}
Beyond recognizing text, many digitization goals require recovering high-level structure from visual layouts, including figures, captions, tables, and charts. Systems for mining PDFs at scale (e.g., PDFFigures 2.0) extract figures, tables, and captions to enable downstream semantic indexing and analysis \cite{clark2016pdffigures2}. A major subarea concerns table understanding: early deep approaches demonstrated end-to-end table detection and cell-structure inference from images without relying on PDF metadata \cite{schreiber2017deepdesrt}. The field then expanded via large benchmarks that pair table images with structured targets. PubTabNet, for example, provides table images aligned with HTML and introduced TEDS, a structure-sensitive similarity metric that better reflects the impact of structural errors than token-only measures \cite{zhong2020pubtabnet}. More recent datasets such as PubTables-1M aim to reduce ground-truth ambiguity and support multiple table tasks (detection, structure, and functional analysis), enabling broader evaluation and training of transformer-based table extraction models \cite{smock2022pubtables1m}. Complementary weakly supervised resources like TableBank increase coverage of table appearance diversity across document sources \cite{li2020tablebank}, while end-to-end table extraction models (e.g., CascadeTabNet) continue to push joint detection+structure recognition \cite{prasad2020cascadetabnet}. Structure recovery also extends to charts and plots. Methods like ChartOCR combine learned keypoint detection with rule-based reconstruction to produce underlying data tables, explicitly privileging controllable intermediate representations \cite{luo2021chartocr}. ChartQA and PlotQA demonstrate that answering real chart/plot questions depends on faithful structure recovery and reasoning, and both highlight persistent gaps when questions require complex logic or out-of-vocabulary numeric answers \cite{masry2022chartqa,methani2020plotqa}. Recent “modality conversion” approaches such as DePlot translate plots into linearized tables and then leverage LLM reasoning, suggesting that reliable conversion to structured representations can be a powerful bridge between perception and downstream analytics \cite{liu2023deplot}. Overall, the literature shows rapid progress in extracting formal structure, but also reveals that small local mistakes (e.g., spanning cells, legend mapping, reading order) can cause large semantic failures motivating interactive correction strategies that can repair and refine structure when fully automated pipelines fall short.

\subsection{Human-in-the-loop correction and instruction-driven editing.}
Human-in-the-loop (HITL) paradigms provide a principled framework for tackling the “last-mile” reliability challenges of document digitization. Work in interactive machine learning emphasizes that rapid feedback cycles, user-visible previews, and careful study of end-user behaviors are central to effective interactive systems \cite{amershi2014power}. Related work further shows that allowing users to contribute interpretable corrections can strengthen trust and satisfaction in interactive ML workflows \cite{10.1145/3490099.3511111}. In end-user programming and data preparation, programming-by-example systems such as FlashFill and FlashExtract show how users can specify intent through examples, with systems synthesizing transformations and actively querying ambiguous cases to converge quickly \cite{gulwani2011flashfill,le2014flashextract}. HCI research on interactive data transformation (e.g., Wrangler) further demonstrates the value of mixed-initiative workflows that combine direct manipulation with inferred operations during iterative refinement \cite{kandel2011wrangler}. More recent HCI work on collaborative editing similarly shows the value of dividing high-level intent specification and low-level operation execution between the user and the system \cite{10.1145/3544548.3580676}. In document processing, post-OCR correction has been widely studied as a complementary stage to OCR, often framed as character-level or seq2seq translation from noisy OCR output to clean text; strategies such as ensemble voting over segmented strings address long-document correction and achieve strong results on multilingual benchmarks \cite{ramirezorta2022postocr,lyu2021neuralocr}. More recently, LLM-centered approaches have been explored for post-OCR correction, leveraging instruction tuning and prompt-based interfaces to detect and repair OCR errors, especially in degraded historical corpora where OCR quality is a major barrier \cite{thomas2024llmpostocr}. Related HCI work also suggests that lightweight natural-language corrections can be an effective mechanism for interactively improving model behavior without requiring extensive relabeling or retraining \cite{10.1145/3654777.3676362}. At the same time, HCI studies of prompt-based system design argue that “prompting is brittle”: while prompts can quickly fix many visible failures, the remaining corner cases are difficult to isolate and can interact in complex ways, motivating tools that provide structure, guardrails, and systematic evaluation of edits \cite{zamfirescu2023herding}. This is consistent with work showing that chaining LLM operations and exposing intermediate steps can improve transparency, controllability, and collaborative use \cite{10.1145/3491102.3517582}. More broadly, recent HCI work on AI-integrated workflows shows that proactive assistance can improve efficiency while also introducing workflow disruption, underscoring the importance of user awareness, contextual cues, and mechanisms that preserve control during interaction \cite{pu2025codellaborator}. Related design studies of AI-powered assistants in notebooks similarly emphasize the value of scoped assistance, disambiguation, and workflow-sensitive support rather than overly intrusive automation \cite{10.1145/3544548.3580940}. Digitization infrastructures for cultural heritage and archival workflows (e.g., systems built around handwritten-text recognition) similarly emphasize that incorporating user corrections can both improve access and enable iterative performance gains over time \cite{seaward2019transkribus}. Finally, large-scale digitization efforts explicitly integrate human validation to overcome bottlenecks in converting images to trustworthy text, reinforcing the view that workflow and interface design can be as critical as model choice \cite{guralnick2024humans}. More broadly, HCI work on provenance-aware LLM interaction suggests that making model contributions visible can help users retain a sense of control and ownership during AI-assisted editing \cite{10.1145/3613904.3641895}. Together, these lines of work motivate instruction-driven document editing interfaces that support localized correction, previewable changes, and propagation of fixes across repeated structures, bridging automated extraction with user-steerable re-shaping toward faithful, editable structured outputs.

\section{Formative Study}

To inform the design of our system, we conducted a formative study with participants who regularly work with scanned documents. We collected responses and conducted an interview with four participants, including student researchers and domain practitioners, and a senior researcher involved in long-term archival digitization initiatives. Participants were asked to describe their typical digitization tasks, workflows, and challenges when working with real-world documents.

Across participants, documents were heterogeneous in structure and quality, including scanned print, handwritten notes, mixed layouts, and complex tables with irregular formatting. The scale of work also varied substantially, ranging from tens of pages per project to collections spanning thousands of pages. Participants reported workflows that combine OCR with substantial manual cleanup, often involving general-purpose tools such as PDF editors, cloud-based word processors, and spreadsheet applications.

In some cases, participants avoided OCR altogether due to low reliability, instead manually extracting relevant information.

Across participants, breakdowns reported in these workflows were dominated by structural errors rather than isolated character recognition mistakes. Participants reported issues such as incorrect reading order, fragmented paragraphs, and misaligned tables, with table reconstruction being particularly error-prone. These structural inconsistencies made verification difficult, as errors were not always immediately visible but instead emerged through inconsistencies in organization. As a result, correcting a single page—especially one containing tables—could take up to 10--20 minutes.

From this perspective, two key requirements emerged. First, digitization systems must support both \textit{data extraction} (e.g., tables, numerical values) and \textit{contextual understanding} (e.g., narratives describing experimental design and methodology). Second, scalability is critical: while manual extraction may be feasible for small datasets, large collections consisting of tens of thousands of documents require automated or semi-automated approaches that can operate efficiently while maintaining accuracy.
At the same time, users expressed concerns about automated systems making unintended changes, particularly in numerically sensitive regions such as tables. To bridge this gap, establishing trust depends on having clear control over the scope of modifications and the ability to review changes before they are applied.

Taken together, these findings suggest that document digitization should be framed as an interactive and iterative refinement process rather than a fully automated pipeline. First, systems must operate at the level of document structure, enabling users to correct and reconstruct tables, layouts, and relationships between elements. Second, systems should reduce redundant effort by allowing users to propagate corrections across similar regions while maintaining explicit control over scope. Additionally, participants reported that beyond repetitive edits, expressing corrections through natural-language prompts alone was often insufficient, as LLMs struggled to accurately interpret localized or structurally complex issues. This highlights the need for interaction mechanisms that combine prompting with more direct and grounded forms of input. Third, digitization outputs must go beyond static text extraction and support transformation into structured, analyzable representations that enable downstream use, including search, aggregation, and scientific analysis. Finally, given the scale and long-term value of many document collections, systems must balance automation with user oversight, ensuring both efficiency and reliability. These insights directly informed the design of our system, which emphasizes human-in-the-loop refinement, structure-aware editing, and controlled propagation of changes to support scalable and trustworthy document digitization workflows.

\section{Document Digitization System}

We present an interactive document digitization system that combines automated extraction with human-guided refinement. The system is designed to transform scanned documents into structured, editable representations while supporting iterative correction and restructuring. Rather than treating digitization as a fully automated pipeline, our approach frames it as a collaborative process in which users and models jointly refine outputs.

\subsection{System Overview}

\begin{figure}[htbp]

  \centering
\includegraphics[width=1.0\linewidth]{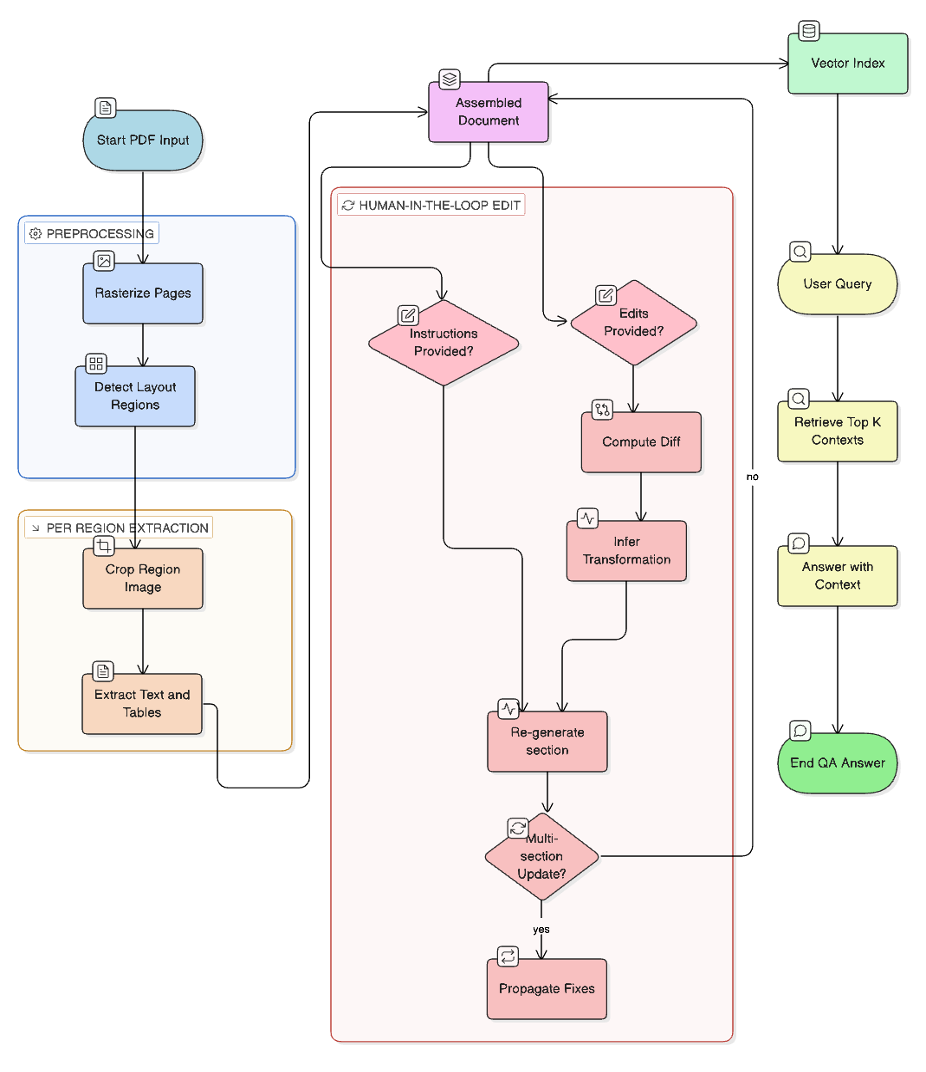}
\caption{\textbf{System workflow.}
    Starting from a PDF document, our pipeline first rasterizes pages and detects layout regions. Each region is then cropped and processed to extract text and tables, which are assembled into a structured document. Users can refine this assembled output through human-in-the-loop interaction by providing direct edits and/or natural-language instructions. When edits are made, the system computes the difference, infers the underlying transformation, re-generates the affected section, and optionally propagates the same fix across multiple related sections. The resulting structured document is also indexed for retrieval, enabling downstream question answering through vector search and context-based response generation.}
    \Description{A workflow diagram showing the system pipeline from PDF input through preprocessing, per-region extraction, human-in-the-loop editing, optional multi-section propagation, and retrieval-based question answering using a vector index.}
  \label{fig:workflow}
\end{figure}

\begin{figure}[htbp]

  \centering

  \includegraphics[width=0.8\linewidth]{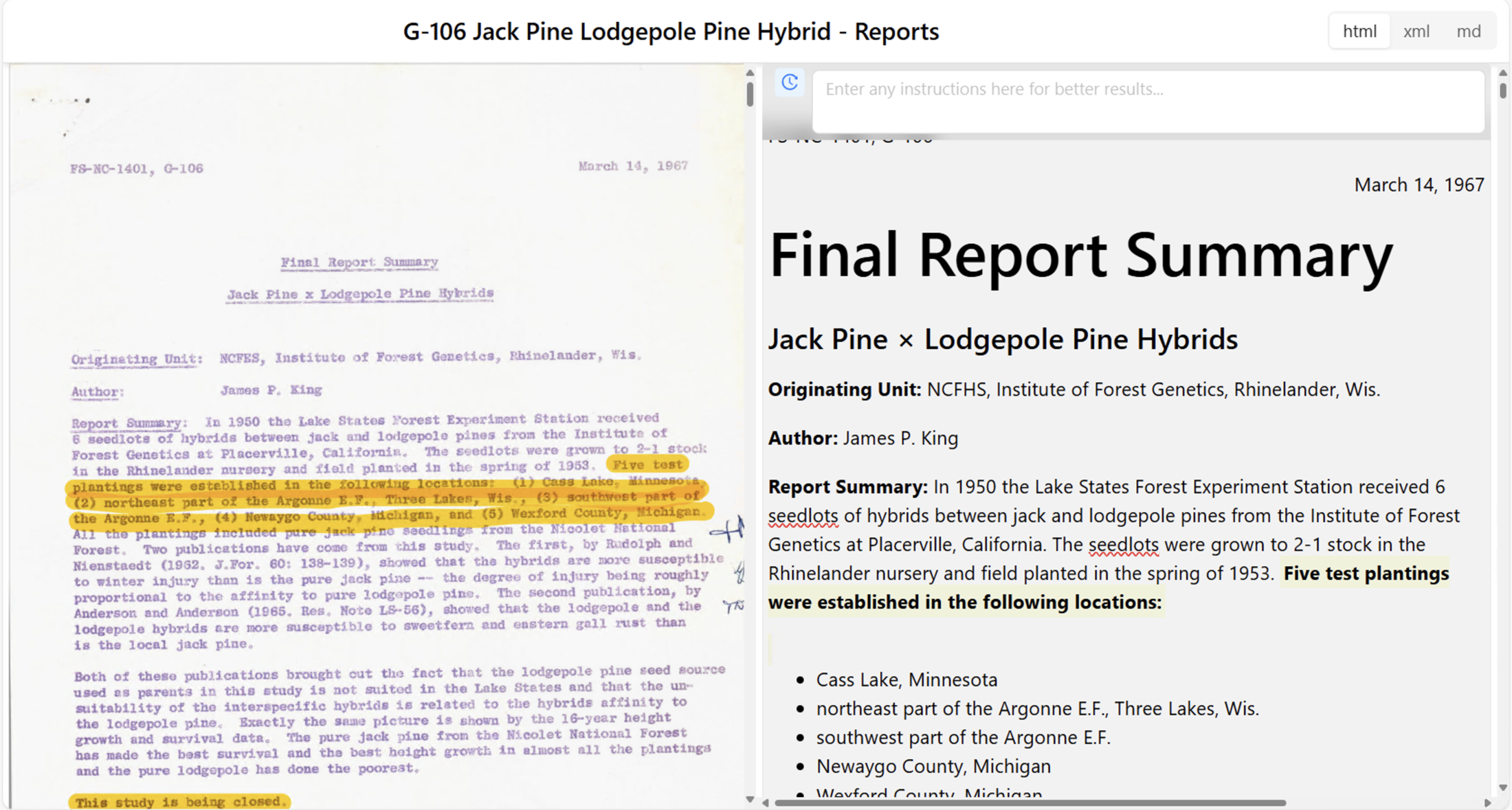}
  \caption{The web-based user interface shown to the user. The left hand panel shows PDF the user uploads to the system. The right hand side shows the extracted version of the output. Users can freely modify the extracted version.}
  \Description{This figure shows the web-based user interface of the system. The left panel displays the original uploaded PDF document, with portions of the text highlighted in yellow. The right panel shows the extracted digital version of the same content, reformatted for editing. Users can interact with this view by directly modifying the extracted version, enabling corrections and adjustments.}
  \label{fig:UI}

\end{figure}

\begin{figure}[htbp]

  \centering

  \includegraphics[width=1\linewidth]{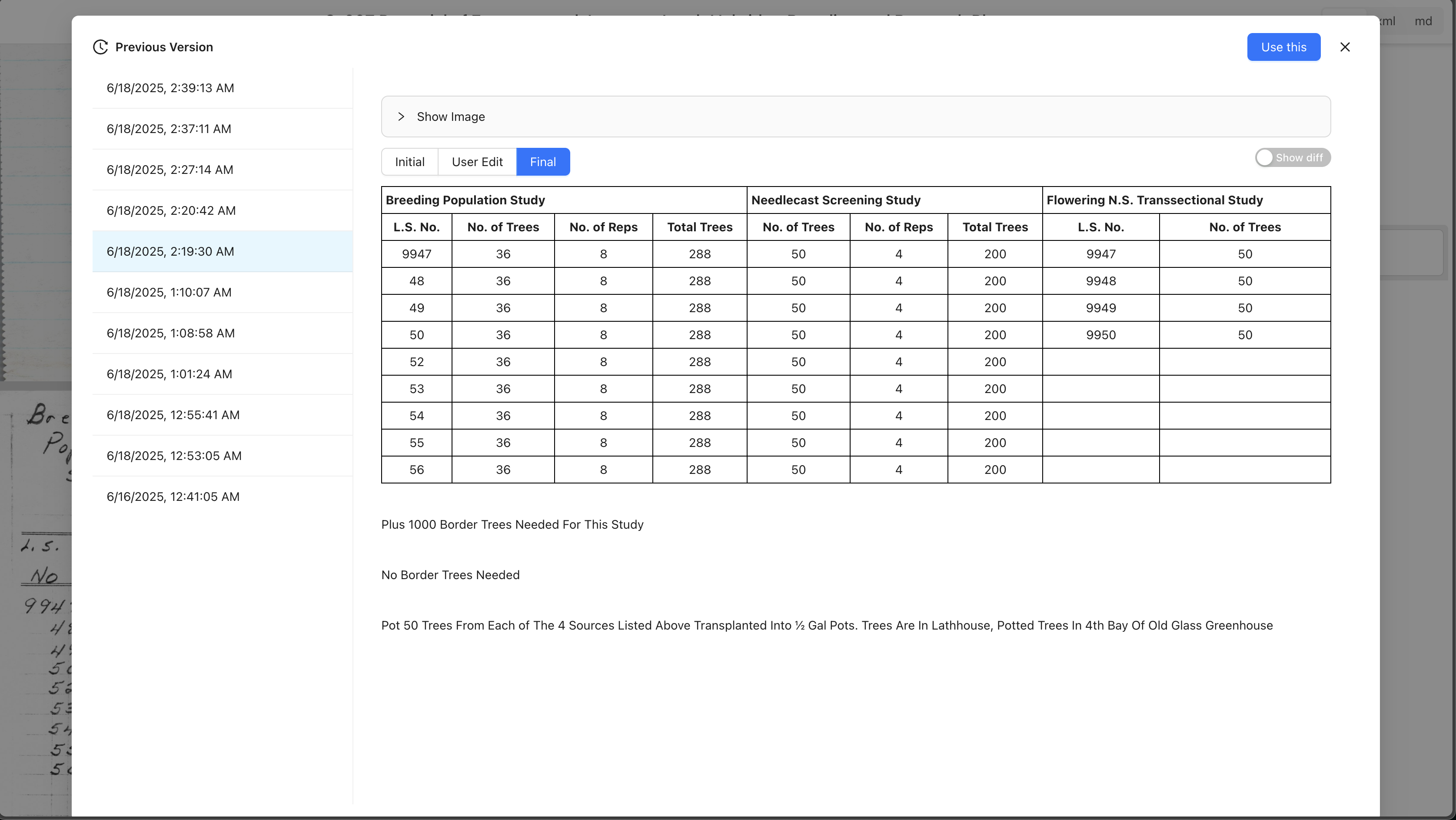}
  \caption{The history of each section is shown in a dialog box. User can go through each interaction, check what was the starting version, what user changes, what instruction user provided and the final output generated by the system. User have the ability to use the specific version that they want to revert back to.}
  \Description{This figure shows the history view of each section displayed in a dialog box. The left panel lists timestamps for different versions, while the right panel shows the corresponding extracted text or table for each version. Users can review the starting version, their edits, the provided instructions, and the final system-generated output. The interface also allows users to select and revert to a specific previous version if desired.}
  \label{fig:History}

\end{figure}

Users interact with the system through a web-based interface that presents the original document alongside its extracted representation (~\autoref{fig:UI}). ~\autoref{fig:workflow} provides an overview of the full pipeline. Starting from a PDF input, the system first performs preprocessing and per-region extraction to assemble a structured document, which is then exposed to users for human-in-the-loop refinement. The front-end is implemented in React, handling document rendering and user interactions, while a Django backend manages document processing, model inference, and data storage. The interface displays each document page as an image aligned with a structured, editable representation generated by the pipeline. This side-by-side view allows users to continuously compare source content with the generated output during correction. Content is organized into layout-based segments (e.g., text blocks, tables, figures), allowing users to focus on localized regions while maintaining consistency across the document. Each segment is directly editable. Users can either modify the extracted content or provide natural-language instructions describing the desired change. These updates are sent to the backend, where they are processed and returned to the interface for re-rendering. After each update, the system presents a comparison between versions and maintains a history of edits (~\autoref{fig:History}), allowing users to inspect, revert, and track modifications over time. This design supports transparency and enables users to retain control over the refinement process.

As shown in \autoref{fig:workflow}, the pipeline consists of three connected stages: document preprocessing and region-level extraction, human-in-the-loop refinement of the assembled document, and optional retrieval-based question answering over the final structured output.

The extraction pipeline corresponds to the left side of ~\autoref{fig:workflow} and converts input documents into structured representations through three stages: layout parsing, text extraction, and structured reconstruction. First, each page is converted into an image and processed by a layout parsing model to identify regions such as text, tables, and figures. The model produces bounding boxes and semantic labels, enabling region-specific processing. Second, each detected region is passed through an OCR engine to extract textual content. While this step produces machine-readable text, it often lacks structural organization and may contain recognition errors. Third, the system uses a large language model (LLM) to reconstruct structured outputs by combining OCR results with visual context through prompt-based conditioning, where layout information and extracted text are jointly provided to the model to guide structured generation. The model generates representations in formats such as HTML or Markdown, preserving layout information when possible. Markdown is used for textual content due to its efficiency, while HTML is used for complex structures such as tables that require richer representation. For figures, image content is extracted but the image is also preserved and added to the extracted document.

\subsection{Layout-Aware Processing}

The layout parsing stage is one of the core component of the pipeline, responsible for detecting and segmenting document pages into semantically meaningful regions such as text blocks, tables, figures, and lists. We formulate layout analysis as an object detection problem and employ a Swin Transformer V2 backbone \cite{liu2022swintransformerv2scaling} combined with a DETR-style\cite{carion2020endtoendobjectdetectiontransformers} detection head to predict bounding boxes and region labels. The model is initially trained on large-scale document layout datasets such as PubLayNet \cite{zhong2019publaynetlargestdatasetdocument} , which provides extensive annotations for common document structures, enabling the model to learn generalizable spatial patterns across diverse document types. To address domain shift in complex historical handwritten documents, we further augment training with synthetic data that mimics historical layouts and noise patterns, improving robustness on degraded and handwritten materials.

The extracted layout information is not only used to segment the document for downstream OCR and LLM processing, but also plays a critical role in guiding the LLM interaction. Specifically, layout regions are used to construct layout-aware, region-specific system prompts, enabling the model to perform in-context learning tailored to different structural elements (e.g., tables versus paragraphs). Furthermore, layout information allows the system to identify structurally similar regions across a document. Since repeated layouts such as tables or templated sections often share consistent structure and recurring errors, user corrections applied to one region can be propagated to others with similar layout characteristics. This enables efficient and consistent refinement across the document, reducing redundant user effort while maintaining structural fidelity.

\subsection{Interactive Refinement and Document Reshaping}

The interactive refinement stage corresponds to the central human-in-the-loop branch in ~\autoref{fig:workflow}. After initial extraction, users iteratively refine the assembled output through direct edits or natural-language instructions. This design supports not only localized correction, but also broader document reshaping, allowing the extracted result to function as an editable intermediate representation rather than a fixed OCR output.

In addition to fixing recognition errors, users can restructure extracted content, reformat tables, transform handwritten notes into cleaner tabular representations, elaborate figures into textual descriptions for downstream semantic search, extract data from visualizations, and generate more structured visual summaries from tabular content. These interactions range from local fixes, such as restoring truncated values or correcting omitted text, to representational changes, such as converting coded values into semantic labels or reorganizing free-form notes into structured tables. Appendix~\ref{app:creative-scenarios} shows representative examples of these interactions, including marked-content enhancement, repeated-text removal, example-guided structural correction, and direct edits without explicit instructions.

When a user provides input, the system prompts the LLM with the original image region, the current extracted content, and any user-provided edits or instructions. For direct edits, the system first computes the difference between the current and modified content, then uses this signal to infer the intended transformation before re-generating the affected section, as illustrated in ~\autoref{fig:workflow}. Grounding the update in both visual and textual context helps address a common limitation of instruction-only editing, where users often struggle to express precise corrections through prompts alone.

Users can also control the scope of modifications, which determines how edits are propagated across the document. Through the interface, they can apply a change only to the current segment, extend it to other segments of the same layout type, or propagate it across the full document. This corresponds to the multi-section update branch in ~\autoref{fig:workflow} and gives the system an explicit signal of which regions should be updated. When propagation is enabled, the system uses layout parsing information to identify structurally similar segments and uses the user’s edit as an in-context example to guide updates in those regions. For each target segment, the prompt includes both the original and corrected source segment, allowing the system to infer the intended transformation and apply it consistently.

\section{Evaluation}
We conducted a within-subjects study to evaluate ReforMe in facilitating iterative document digitization. Comparing our system against a standard LLM-based workflow, we investigated how our design supports three core capabilities: localized correction of extraction errors, automated propagation of changes across similar content, and high-level reshaping of document outputs.

\subsection{Participants} 12 participants were recruited through email lists and word of mouth (10 male, 2 female; age range 18--32). Before beginning the study, participants completed a short intake questionnaire about their prior experience with digitization workflows, OCR tools, and AI chat systems. Overall, the sample reflected greater familiarity with AI chat tools than with OCR-specific workflows: participants reported moderate frequency of working with digitization/OCR systems ($M=2.83$ on a 6-point scale), relatively limited familiarity with OCR tools ($M=2.17$ on a 5-point scale), and higher familiarity with AI chat tools for document-related tasks ($M=4.00$ on a 5-point scale). They also generally viewed accurate document extraction as important to their work ($M=3.33$ on a 5-point scale) and reported feeling prepared to complete the study tasks ($M=5.83$ on a 7-point scale).

\subsection{Tasks}

The controlled study consisted of three task types, each designed to evaluate a different aspect of the workflow. In all three tasks, participants were shown the original document, the extracted result produced by the system, and the target outcome they were asked to obtain.

\begin{enumerate}
    \item \textbf{Task 1: Localized Error Correction.} Participants worked with a table in which values in the first column had been truncated during extraction. For example, the second row correctly contained the value \textit{11977}, but subsequent rows contained shortened values such as \textit{78}, which should instead be interpreted as \textit{11978}. The goal of the task was to restore the missing leading digits and produce the full non-truncated table, as illustrated in \autoref{fig:task1_correction}. This task focused on localized correction, where the required fix was conceptually simple but had to be applied accurately within the existing table structure. It was intended to evaluate whether direct modification, combined with lightweight instruction, supports precise correction of local extraction errors.

    \begin{figure}[t]
        \centering
        \includegraphics[width=\columnwidth]{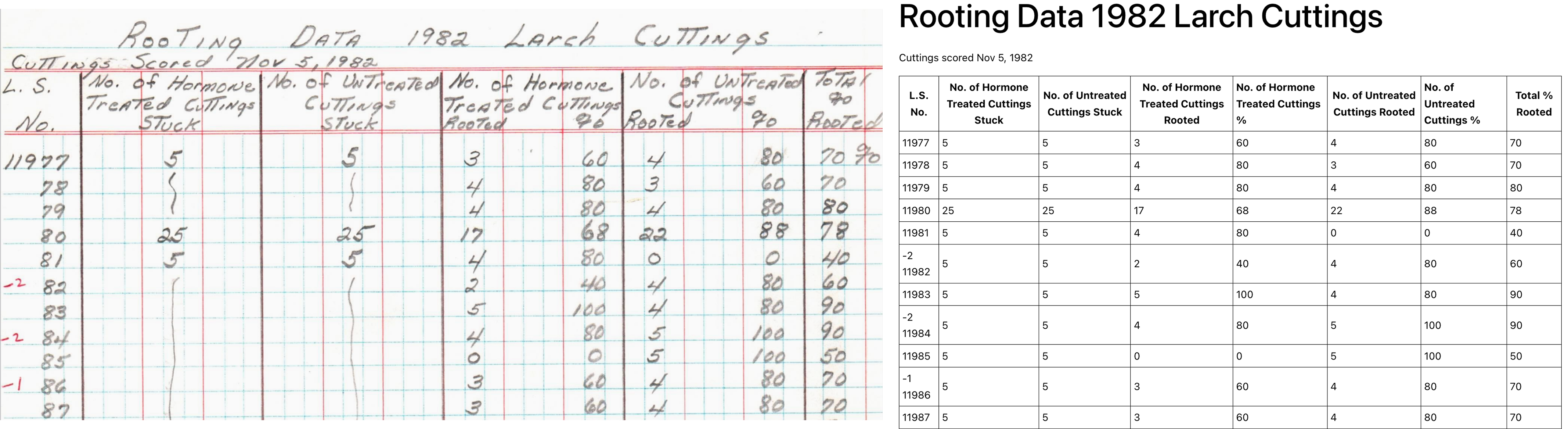}
        \caption{Task~1. The left image shows the original document and the right image shows the target result participants were asked to obtain. The task required restoring truncated values in the first column of the extracted table.}
        \label{fig:task1_correction}
    \end{figure}

    \item \textbf{Task 2: Propagation Across Repeated Structure.} Participants worked with a multi-page table in which checkbox information from the source document had not been included in the generated output. In the original pages, some entries were marked with a checkbox indicating that the corresponding tree was \textit{dead}, but this status was missing from the extracted table. Participants were asked to modify the output so that entries corresponding to checked boxes were labeled as \textit{dead}, and to apply this same fix consistently across all relevant pages. As shown in \autoref{fig:task2_propagation}, the same issue occurred across seven pages. This task was designed to evaluate whether the workflow supports change propagation more effectively than a standard LLM-based workflow, especially when the same correction must be applied repeatedly across structurally similar regions.

    \begin{figure}[t]
        \centering
        \includegraphics[width=\columnwidth]{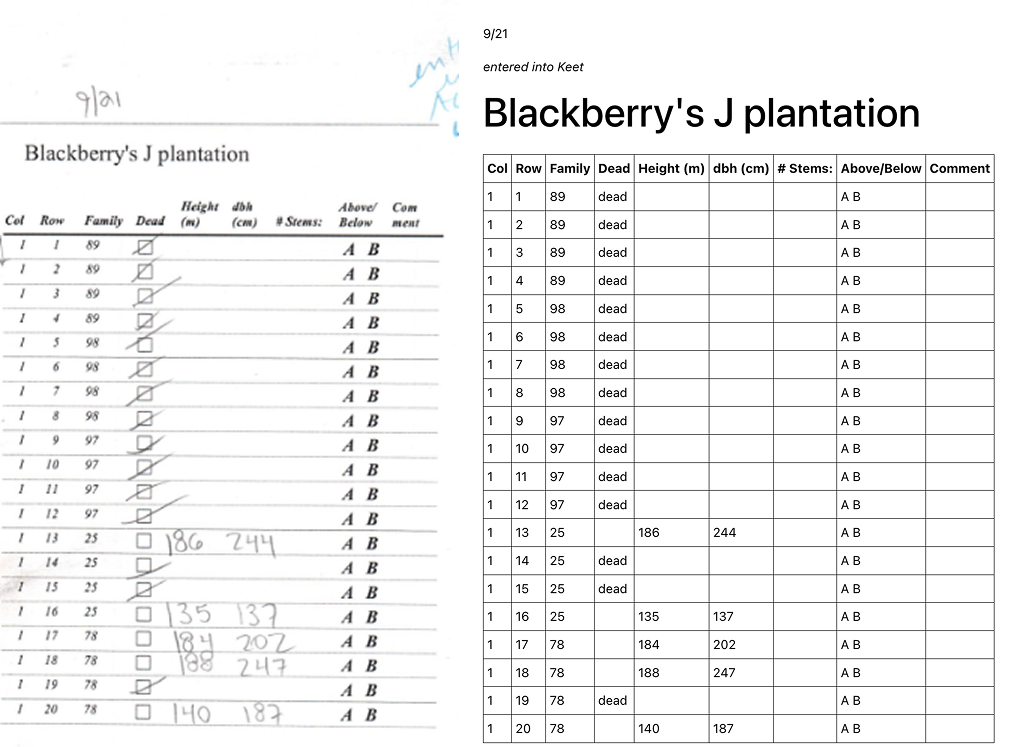}
        \caption{Task~2. The left image shows the original document and the right image shows the target result participants were asked to obtain. The same correction-adding \textit{dead} where the checkbox indicates tree death-needed to be applied consistently across seven pages.}
        \label{fig:task2_propagation}
    \end{figure}

    \item \textbf{Task 3: Reshaping Using a Legend.} Participants worked with a generated table whose cells contained symbolic or numeric codes copied directly from the source document. Their task was to convert these codes into a more interpretable representation using the legend provided in the original document. Specifically, they were asked to transform the values according to the mapping \textit{1 = Good, 2 = Fair, 3 = Poor, 4 = Dead, X = Missing}, and then apply the corresponding visual encoding \textit{Good = green, Fair = light green, Poor = yellow, Dead = red, Missing = gray}. The goal was therefore not only to replace the codes with semantic labels, but also to visually restructure the table through color coding, as shown in \autoref{fig:task3_reshape}. Participants could use multiple prompts if needed. This task was designed to evaluate higher-level reshaping, which requires a more complex transformation than straightforward error correction and is often cumbersome to specify and verify in a conventional chat-based workflow.

    \begin{figure}[t]
        \centering
        \includegraphics[width=\columnwidth]{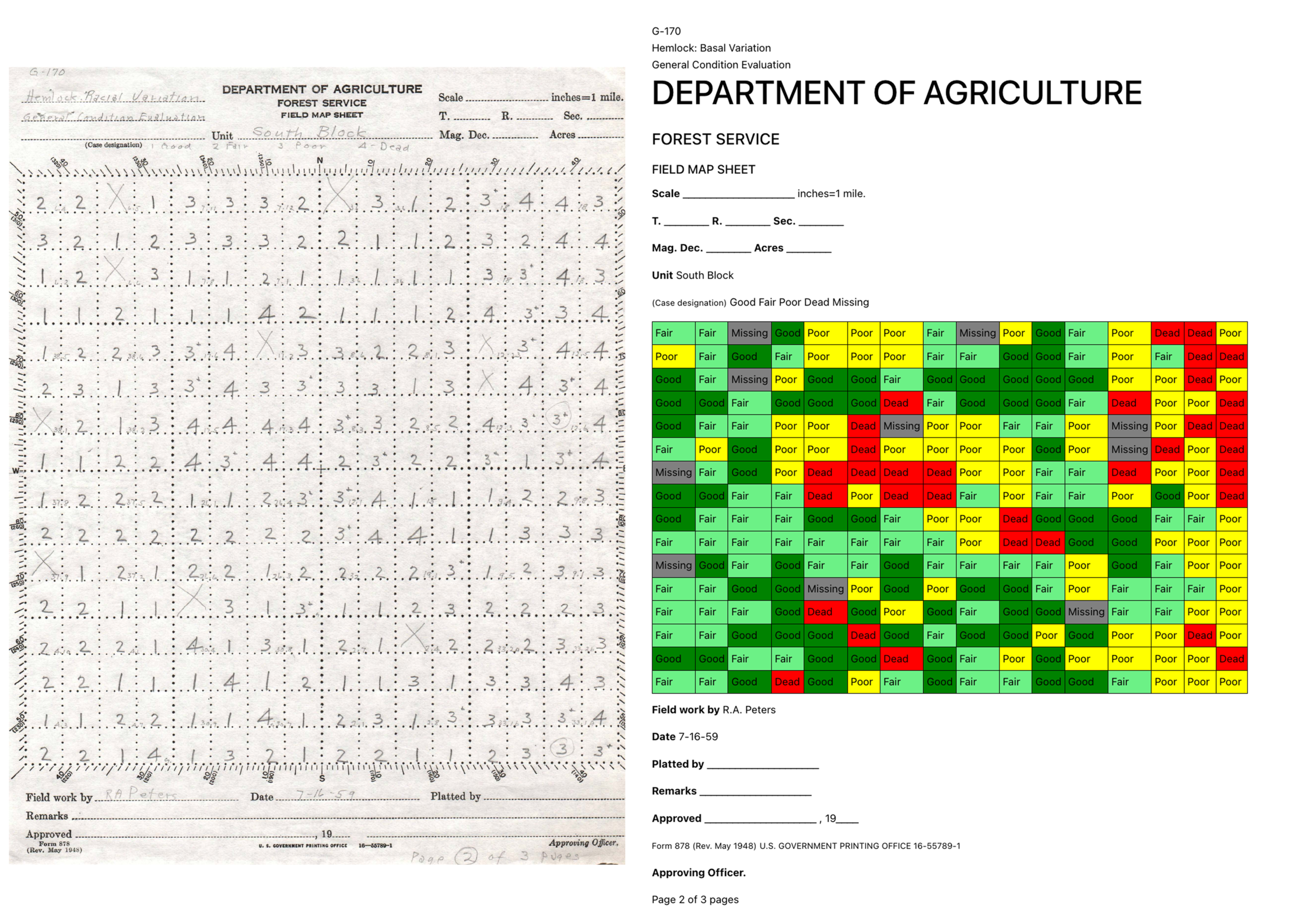}
        \caption{Task~3. The left image shows the original document and the right image shows the target result participants were asked to obtain. The task required converting numeric or symbolic codes into semantic labels based on the legend and applying corresponding color encodings.}
        \label{fig:task3_reshape}
    \end{figure}
\end{enumerate}

Because these tasks required participants to inspect outputs, correct local errors, propagate repeated edits, and verify system behavior, they also provided a basis for assessing the usefulness of interaction features such as direct edits, propagation, preview, and history.

\subsection{Conditions}
The study compared two workflows. In the baseline condition, participants used a locally hosted LLM chat interface (LibreChat \cite{librechat})  to perform the tasks. In the second condition, participants used our web-based system. We intentionally distinguished between \emph{initial extraction quality} and \emph{interactive refinement}, since our contribution lies not only in the generated output, but also in the mechanisms that support inspection, correction, propagation, and restructuring.

\subsection{Procedure}
Each participant completed all tasks with both workflows, allowing us to compare the two approaches on the same refinement problems. Each session began with consent, the background questionnaire, and a brief introduction to the interface, followed by a short demo task. Participants then evaluated the initial extraction outputs produced by each workflow and completed the three study tasks. Afterward, they completed post-task and post-study questionnaires and provided open-ended feedback.

\subsection{Measures}
We collected 7-point ratings together with open-ended responses. The quantitative measures covered three levels. First, for the initial extraction stage, participants rated each workflow on effectiveness, effort, and confidence. Second, for each task, participants rated both workflows on completion effectiveness and effort, together with a task-specific comparative item aligned with the task objective: ease of applying corrections in Task~1, ease of making the same change across many pages in Task~2, and support for structural or reshaping changes in Task~3. Third, for our system, we collected system-specific ratings on usability and interaction quality, including ease of use, instruction following, edit following, repetitive-work reduction, propagation usefulness, perceived control, history, preview, and support for intermediate verification. We also included workload measures adapted from NASA-TLX \cite{HART1988139}, specifically mental demand and frustration. Participants additionally rated overall confidence, trust, and likelihood of future use for both workflows, and provided open-ended comments about the most useful aspect of the system and the most important improvement they would suggest.

\subsection{Analysis} We analyze the controlled study at two levels. First, we compare the two workflows on initial extraction and task-specific ratings to assess their relative effectiveness and effort. Second, we examine system-specific ratings and open-ended responses to better understand how participants experienced propagation, transparency, and control across the three task types.

\section{Results}

\begin{figure}[t]
    \centering
    \includegraphics[width=\columnwidth]{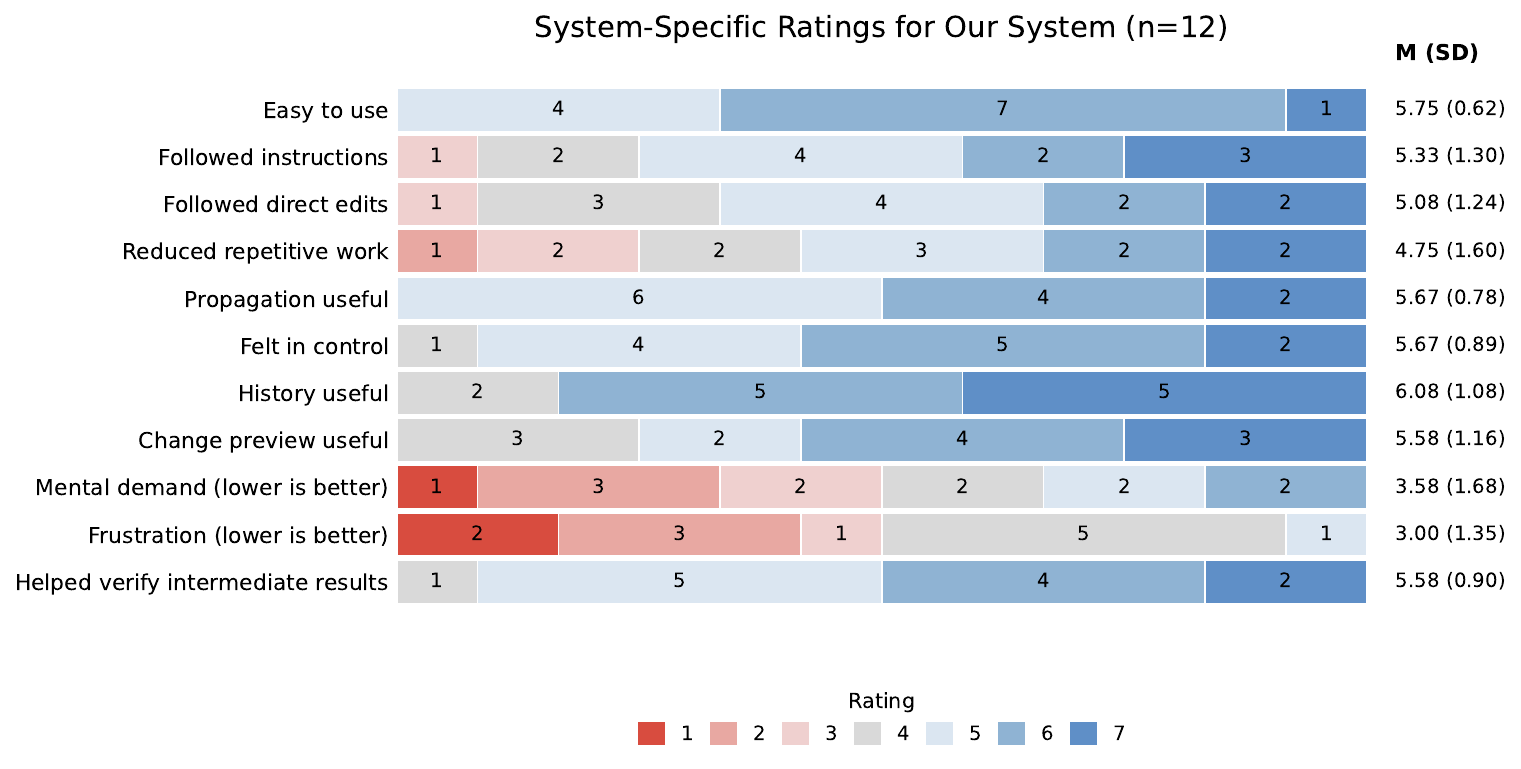}
    \caption{Distribution of system-specific ratings for our system in the controlled study ($n=12$). Bars show response counts on 7-point scales; the rightmost columns report mean (SD) and favorable responses. Favorable responses correspond to ratings of 6--7 for positive items and 1--2 for workload items.}
    \label{fig:controlled-system-ratings}
\end{figure}

We analyzed the study data from 12 participants. We use paired Wilcoxon signed-rank tests to assess whether within-participant rating differences between the two workflows are systematic across matched tasks. Because we conduct 15 paired comparisons, we also report Benjamini--Hochberg false-discovery-rate (FDR) corrected $q$-values to control for multiple testing. The comparative results are summarized in Table~\ref{tab:controlled-comparison} and visualized in Figure~\ref{fig:controlled-system-ratings}. Overall, our system showed the strongest advantages in initial extraction, repeated-change propagation, higher-level reshaping, and overall confidence.

\paragraph{Participants favored our system at initial extraction}

Even before interactive refinement, that is, the first generated output shown before any user edits or prompts, participants generally preferred the initial extraction results produced by our system over those produced by the standard LLM-based workflow. Our system received higher ratings for extraction effectiveness (our system: $M=5.92$, $SD=0.67$; Chat: $M=4.33$, $SD=1.30$),
and this difference was statistically reliable ($p=.0068$, $q=.0293$).

There were no significant differences in the efforts between the conditions our system during initial extraction (our system: $M=2.67$, $SD=1.50$; Chat: $M=4.08$, $SD=1.38$) ($p=.0645$, $q=.0906$). Confidence showed the same direction, increasing from $M=3.08$, $SD=1.51$ for the chat workflow to $M=4.42$, $SD=1.24$ for our system, although this difference did not reach statistical significance ($p=.0664$, $q=.0906$).

Participants' comments suggest that this early advantage stemmed less from perfect extraction and more from producing a better starting point for downstream refinement. One participant noted that ``the extraction was quite smooth without having to constantly check if it was doing everything correctly,'' whereas the chat workflow ``requires more instructions.'' Another described our system as producing ``a better-organized output,'' while others emphasized that it ``can get same structure all at once'' and ``creates an html that is readable, while the standard ChatGPT gives unstructured results.'' Participants also noted that the system worked better on complex formats and provided more ways to control the generated result. Taken together, these responses suggest that our system's initial advantage lies in producing a more usable and inspectable intermediate representation rather than eliminating the need for correction altogether.

\paragraph{Localized correction benefited from direct editing and prompting}
For Task~1, which focused on localized error correction, the pattern still favored our system, but the advantage was more modest than in other parts of the study. Our system was rated higher on completion effectiveness
Participants also found our system easier for applying corrections: the mean rating increased from $M=4.33$ for the chat workflow to $M=5.08$ for our system.
This improved experience was mirrored by a reduction in perceived effort (our system: $M=3.92$, $SD=1.24$; Chat: $M=4.25$, $SD=0.87$). However, none of the Task~1 comparisons reached statistical significance.

The qualitative responses suggest that our system was most helpful when participants could combine direct demonstration with instruction. One participant explained, ``I could change a few values myself and then tell it I want the rest done in this same manner,'' which made the refinement process feel more controllable and predictable. Other participants similarly highlighted ``the ease of giving prompts and the intuitiveness,'' suggesting that localized correction benefited from a workflow in which users could quickly demonstrate a pattern and refine it with lightweight prompting. At the same time, the comments reveal that localized correction still depends on reliable structural reconstruction. This pattern suggests that our system improved the expression of local corrections, but that precise table alignment remained a bottleneck when those changes had to be integrated back into structured content.

\paragraph{Change propagation reduced effort and improved repeated-change editing}

Task~2 examined whether our system helped participants use change propagation across repeated structure.
Our system showed a significant advantage on the repeated-change item, increasing from $M=2.92$ ($SD=1.73$) for the chat workflow to $M=4.83$ ($SD=1.34$) for our system ($p=.0156$, $q=.0335$). The effort difference also favored our system, with mean effort ratings decreasing from $M=4.92$ ($SD=1.73$) for the chat workflow to $M=3.42$ ($SD=1.51$) for our system, and this difference remained significant after correction ($p=.0293$, $q=.0488$).

System-specific ratings further suggest that participants generally found change propagation useful ($M=5.67$, $SD=0.78$), but ratings for whether our system reduced repetitive correction work were more mixed ($M=4.75$, $SD=1.60$). The comments suggest that participants distinguished clearly between \emph{effort savings} and \emph{reliability}. One participant wrote, ``It added the word `dead' to some entries, but missed out on some other. It did the task, the effort required was minimal, but the accuracy wasn't good.'' At the same time, other comments were strongly positive: one participant noted that the system could ``replace the change on the same element on all documents, and it did it correctly and perfectly.'' This indicates that change propagation was already valuable as a labor-saving mechanism, even though missed or inconsistent updates still limited users' willingness to rely on it uncritically in all cases.

\paragraph{Our system showed clear advantages on reshaping and higher-level transformation}

In Task~3, which required higher-level reshaping rather than straightforward correction, our system was rated substantially higher on completion effectiveness, increasing from $M=2.92$ ($SD=2.23$) for the chat workflow to $M=6.25$ ($SD=0.97$) for our system; this difference was statistically significant ($p=.0039$, $q=.0293$). Our system also required less effort, with mean effort ratings decreasing from $M=4.92$ ($SD=2.11$) for the chat workflow to $M=2.50$ ($SD=1.17$) for our system ($p=.0078$, $q=.0293$).

Participants' comments reinforce that reshaping was one of the system's clearest strengths. One participant described the color-based transformation support as ``awesome,'' while another highlighted ``the comparison system between the upload document and table generated'' together with the difference view after prompting a correction. Other responses similarly emphasized that the system ``provide[s] more similar result when the document [is] in table format,'' ``show[s] what content are changed,'' and ``provide[s] correct color under the instruction.'' The magnitude and consistency of these results suggest that our system's primary advantage is not merely making local edits faster, but supporting transformations that are cumbersome to specify, execute, and verify in a general-purpose LLM workflow.

\paragraph{Participants valued control, transparency, and reversibility}

Participants reported substantially greater confidence in our system than in the chat workflow (our system: $M=5.08$, $SD=1.16$; Chat: $M=3.08$, $SD=1.51$), and this difference was statistically reliable ($p=.0098$, $q=.0293$). Our system also scored higher on overall trust (our system: $M=5.33$, $SD=1.07$; Chat: $M=3.33$, $SD=1.67$) and likelihood of real-world use (our system: $M=5.67$, $SD=0.78$; Chat: $M=3.67$, $SD=1.97$). Both differences remained significant after FDR correction (trust: $p=.0234$, $q=.0439$; likelihood of use: $p=.0156$, $q=.0335$).

Regarding system usability, participants rated our system as easy to use ($M=5.75$, $SD=0.62$), reported a strong sense of control ($M=5.67$, $SD=0.89$), and gave particularly high ratings to the history feature ($M=6.08$, $SD=1.08$) and the change preview feature ($M=5.58$, $SD=1.16$). These features also appeared prominently in the semi-structured interviews. One participant described ``showing history \& moving back to previous output'' as the most useful aspect of the system, while another highlighted the visual comparison support as especially helpful for inspecting changes. Additional comments reinforced this pattern, describing the visual diffs feature as ``probably the best,'' noting that ``the continuity between the cells and the document were useful,'' and pointing out that automatic document splitting was helpful because, in a traditional chat workflow, the user would otherwise need to manually split the PDF to avoid running out of context. In other words, participants' positive judgments were not driven only by output quality; they also reflected the interface's support for inspection, reversibility, and iterative verification.

At the same time, the comments reveal limits that remain important for deployment. Participants reported that that larger files could take longer to process. Others wanted the ability to provide instructions before initial processing or to preserve more local context across edits. These observations suggest that further improvements should address not only model performance but also the continuity of the interaction workflow during repeated refinement.

\begin{table*}[t]
\centering
\small
\caption{Comparative ratings in the controlled study ($n=12$). For effectiveness, confidence, trust, and usability-related items, higher values are better. For effort items, lower values are better. The final two columns report the number and percentage of participants whose rating favored our system, and the paired Wilcoxon signed-rank test result with Benjamini--Hochberg FDR correction across the 15 paired comparisons.}
\label{tab:controlled-comparison}
\begin{tabular}{p{4.35cm}p{2.1cm}p{2.1cm}p{2.45cm}p{2.35cm}}
\toprule
\textbf{Measure} & \textbf{Chat} & \textbf{Our system} & \textbf{Our system favored} & \textbf{$p$ ($q$)} \\
\midrule
\multicolumn{5}{l}{\textit{Initial extraction}} \\
Effectiveness & 4.33 (1.30) & 5.92 (0.67) & 10/12 (83.3\%) & .0068 (.0293) \\
Effort (lower is better) & 4.08 (1.38) & 3.00 (1.76) & 7/12 (58.3\%) & .1660 (.2264) \\
Confidence & 3.08 (1.51) & 4.42 (1.24) & 7/12 (58.3\%) & .0664 (.0996) \\
\addlinespace[2pt]
\multicolumn{5}{l}{\textit{Task 1: Localized error correction}} \\
Completion effectiveness & 5.08 (2.02) & 5.50 (1.31) & 6/12 (50.0\%) & .5234 (.5234) \\
Effort (lower is better) & 4.25 (0.87) & 3.92 (1.24) & 6/12 (50.0\%) & .4746 (.5085) \\
Ease of applying corrections & 4.33 (1.56) & 5.08 (1.62) & 8/12 (66.7\%) & .3799 (.4383) \\
\addlinespace[2pt]
\multicolumn{5}{l}{\textit{Task 2: Propagation across repeated structure}} \\
Completion effectiveness & 3.92 (1.98) & 5.08 (1.56) & 7/12 (58.3\%) & .1836 (.2295) \\
Effort (lower is better) & 4.92 (1.73) & 3.42 (1.51) & 9/12 (75.0\%) & .0293 (.0488) \\
Ease of applying same change across many pages & 2.92 (1.73) & 4.83 (1.34) & 9/12 (75.0\%) & .0156 (.0335) \\
\addlinespace[2pt]
\multicolumn{5}{l}{\textit{Task 3: Reshaping}} \\
Completion effectiveness & 2.92 (2.23) & 6.25 (0.97) & 9/12 (75.0\%) & .0039 (.0293) \\
Effort (lower is better) & 4.92 (2.11) & 2.50 (1.17) & 8/12 (66.7\%) & .0078 (.0293) \\
Support for structural/reshaping changes & 2.75 (2.01) & 6.00 (1.04) & 10/12 (83.3\%) & .0020 (.0293) \\
\addlinespace[2pt]
\multicolumn{5}{l}{\textit{Overall perceptions}} \\
Overall confidence & 3.08 (1.51) & 5.08 (1.16) & 9/12 (75.0\%) & .0098 (.0293) \\
Overall trust & 3.33 (1.67) & 5.33 (1.07) & 8/12 (66.7\%) & .0234 (.0439) \\
Likelihood of real-world use & 3.67 (1.97) & 5.67 (0.78) & 8/12 (66.7\%) & .0156 (.0335) \\
\bottomrule
\end{tabular}
\end{table*}

\section{Discussion, Limitations and Future Work}

Our results highlight both the promise and the current boundaries of interactive document digitization. In the controlled study, our system showed the strongest gains in initial extraction, repeated-change editing, higher-level reshaping, and overall confidence. At the same time, the study points to areas where the workflow can be strengthened further.

\paragraph{Preserving User Intent Across Iterations}

Preserving user intent across successive edits. In the controlled study, participants valued the ability to demonstrate a change and ask the system to continue it, especially for repeated-change tasks. At the same time, some propagated updates were incomplete or inconsistently applied. Future systems would benefit from stronger representations of transformation intent across turns, rather than treating each update as an isolated revision.

\paragraph{Transparency, Verification, and Trust}

The results also point to the importance of transparency in supporting trust. Participants especially valued the history and change-preview features, and several highlighted visual comparison and the ability to return to earlier states as among the most useful aspects of the system. These features helped make refinement more understandable and controllable, but users still needed to verify outputs carefully when prompts were interpreted imperfectly or propagated changes were only partially successful. This suggests an opportunity to go beyond after-the-fact inspection and provide stronger support for confirming whether an intended transformation has been applied correctly.

\paragraph{Generalizability Across Document Domains}

Another limitation concerns generalizability. The evaluation was grounded in archival documents that include irregular tables, handwritten annotations, shorthand, and heterogeneous layouts. These materials are well suited for demonstrating the value of interactive refinement, but some observed strengths and weaknesses may reflect the particular demands of such archives. More broadly, degraded scans, dense content, and visually ambiguous source regions remain challenging not only for OCR, but also for downstream LLM-based refinement. Future work should examine how well the workflow transfers to other archival domains with different structural conventions.

\paragraph{Future Directions}

These findings point to several directions for future work. At the modeling level, an important next step is improving structure-aware refinement, especially for tables and other layouts where small positional errors can substantially affect meaning. This includes better support for cell-level alignment, stronger handling of degraded scans, and mechanisms for preserving local structural context during iterative updates. Another promising direction is improving how the system infers and applies the logic behind a user's demonstrated correction.

At the interaction level, participants wanted more control over when and how instructions are applied, including the ability to provide guidance earlier in the process, preserve more context across edits, and target changes at finer levels of granularity. Richer scoping controls, clearer previews of propagated updates, and more flexible combinations of direct edits and natural-language instructions could further strengthen the workflow. Participants also noted practical usability issues such as longer processing times for larger files and difficulty locating relevant interface elements during extended use, suggesting that responsiveness and interface continuity remain important design considerations.

Finally, future evaluations should broaden both scope and duration. Studying the workflow across additional domains and over longer periods of use would help clarify how prompting strategies evolve, how trust develops with repeated interaction, and how features such as history, preview, and propagation support sustained refinement in practice. More broadly, our findings suggest that document digitization systems are most useful when treated not only as extraction tools, but as interactive environments for restructuring, verification, and reuse.

\section{Conclusion}

We presented ReforMe, an interactive document digitization system that reframes digitization from a one-shot pipeline into a collaborative refinement workflow. By combining layout parsing, OCR, and LLM-based reconstruction with direct edits, natural-language instructions, change propagation, and reversible refinement, ReforMe supports not only error correction but also higher-level reshaping of extracted content. Our findings show that this interaction model provides clear benefits over a standard LLM-based workflow, particularly for initial extraction, repeated-change editing, higher-level reshaping, and overall confidence. More broadly, this work suggests that document digitization systems are most effective when designed not only as extraction engines, but as interactive environments for inspection, correction, restructuring, and verification.

\bibliographystyle{ACM-Reference-Format}
\bibliography{sample-base}

\appendix
\section*{Appendix}
\section{Creative Interaction Scenarios}
\label{app:creative-scenarios}

From the interaction logs, we identified a wide range of ways users creatively modified extracted content. In this appendix, we highlight four representative cases that illustrate the breadth of our system's capabilities: (1) recognizing and enhancing marked content, (2) removing page numbers and repeated text, (3) example-guided structural correction, and (4) direct edits without instructions.

\subsection{Recognizing and Enhancing Marked Content}

\begin{figure}[htbp]
  \centering
  \includegraphics[width=0.6\linewidth]{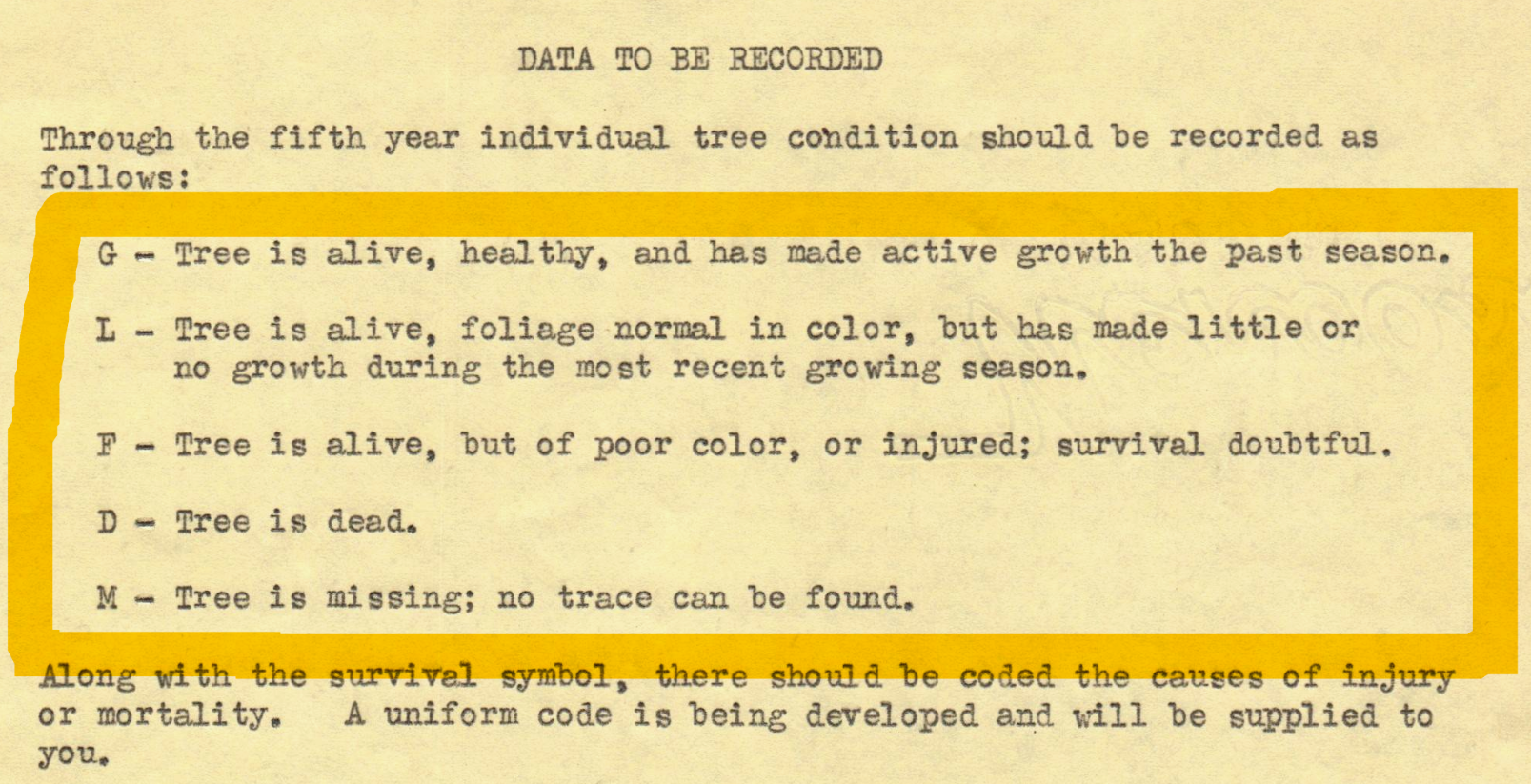}
  \caption{Uploaded document illustrating a case of recognizing and enhancing marked content. The image shows a rectangular section in the middle highlighted with a yellow marker, drawing attention to a list of condition codes for trees.}
  \label{fig:section3-1}
\end{figure}

\begin{figure}[htbp]
  \centering
  \includegraphics[width=1.0\linewidth]{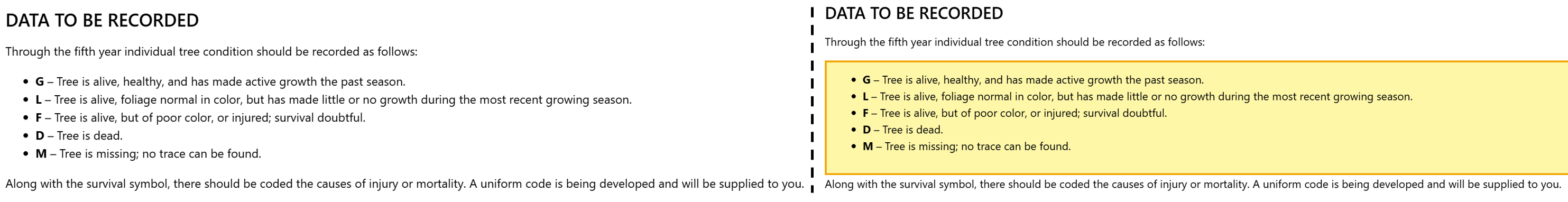}
  \caption{Comparison of the system's recognition and enhanced extraction of the highlighted rectangular section. The left panel shows the initial extraction, where the highlighted section has no special emphasis. The right panel shows the enhanced version generated after the user instruction \textit{``Can you see the mark on the document? This means it is important, enhance the content inside.''} The system recognizes the highlighted region and applies structured formatting, placing the section in a shaded box for emphasis.}
  \label{fig:section3-2}
\end{figure}

One representative interaction involved a manually highlighted block that denoted important information in the document, shown in \autoref{fig:section3-1}. The user provided the instruction \textit{``Can you see the mark on the document? This means it is important, enhance the content inside.''} In response, our system recognized the marked area and enhanced its presentation by applying emphasis. The comparison between the original extraction and the enhanced version is shown in \autoref{fig:section3-2}.

\begin{figure}[htbp]
  \centering
  \includegraphics[width=0.6\linewidth]{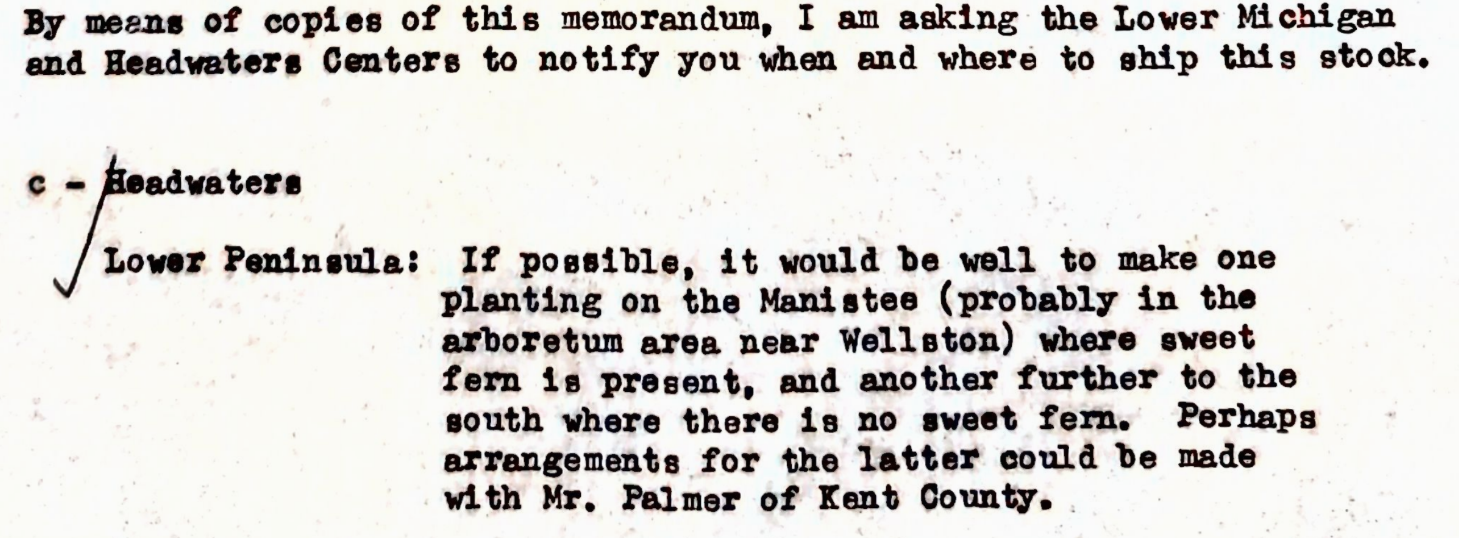}
  \caption{Uploaded document illustrating another case of recognizing and enhancing marked content. The image shows a check mark symbol placed in front of the line beginning with \textit{``c -- Headwaters''}, highlighting the paragraph as important.}
  \label{fig:section3-3}
\end{figure}

\begin{figure}[htbp]
  \centering
  \includegraphics[width=1.0\linewidth]{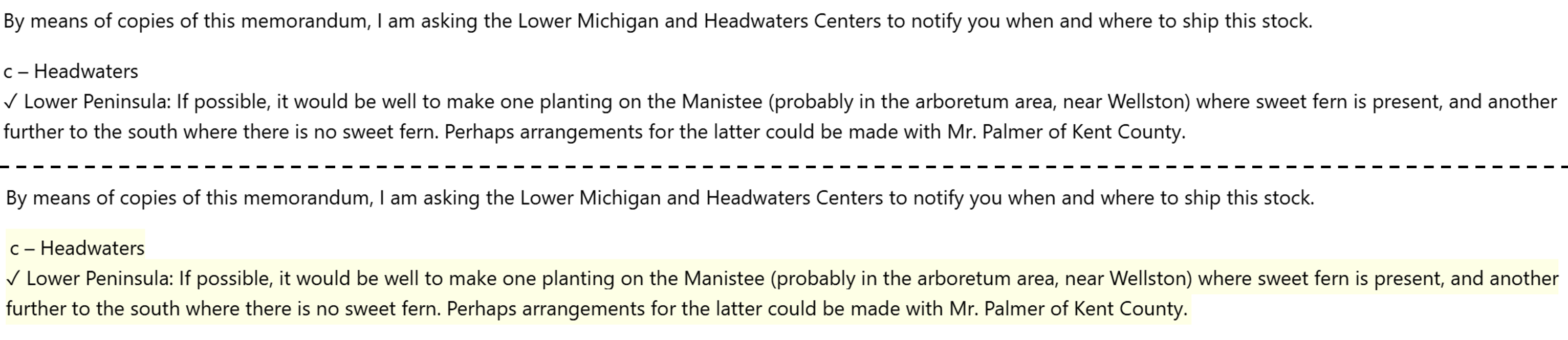}
  \caption{Comparison of the system's recognition and enhanced extraction of special symbols. The top panel shows the paragraph marked with a check symbol, where the annotation is transcribed without emphasis. The bottom panel shows the improved output after the instruction \textit{``There is a check inside this document, which means that the paragraph is important, can you mark the paragraph out,''} where the system recognizes the check mark and highlights the corresponding paragraph.}
  \label{fig:section3-4}
\end{figure}

Beyond recognizing highlights made with a marker or highlighter, the system can also detect check marks and similar annotation symbols in documents. In another case, a user noticed a check symbol in front of a paragraph, indicating that the content was important, as shown in \autoref{fig:section3-3}. The user then provided the instruction \textit{``There is a check inside this document, which means that the paragraph is important, can you mark the paragraph out.''} The system automatically recognized the associated content and emphasized it in the digitized version, as shown in \autoref{fig:section3-4}.

\subsection{Removing Page Numbers and Repeated Text}

\begin{figure}[htbp]
  \centering
  \includegraphics[width=1.0\linewidth]{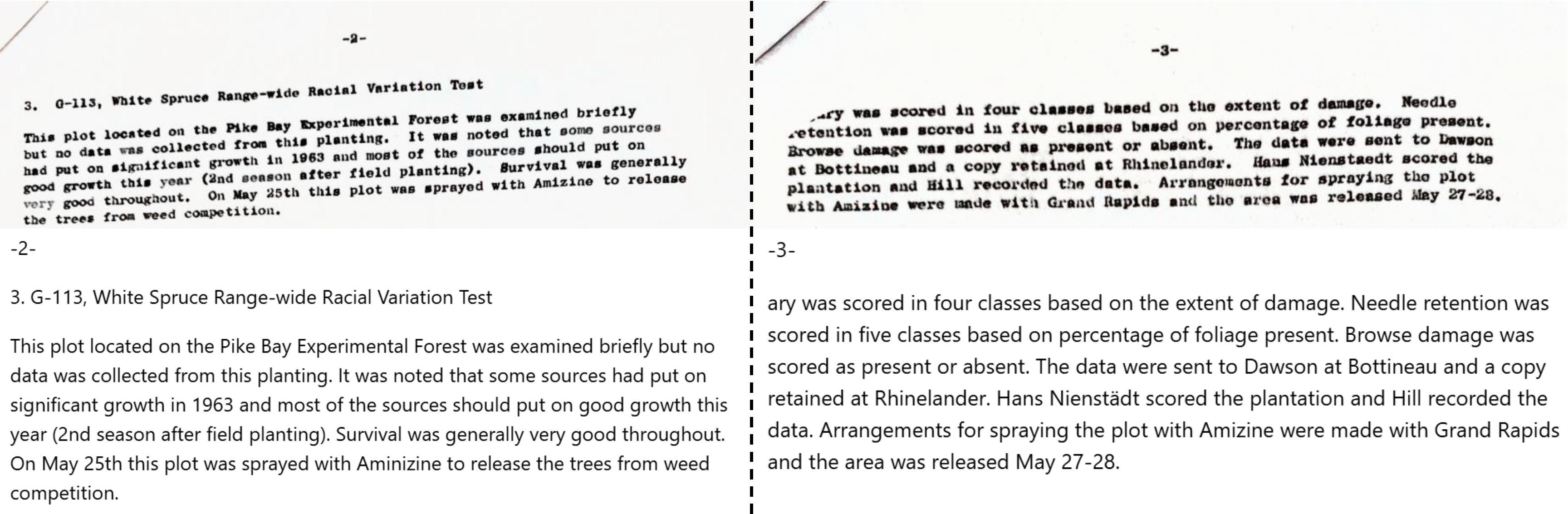}
  \caption{Initial state of the original document and corresponding extracted text for page number removal. The original pages (top) contain page numbers such as ``--2--'' and ``--3--'' embedded within the text. In the extracted version (bottom), these numbers appear as separate lines.}
  \label{fig:section4-1}
\end{figure}

\begin{figure}[htbp]
  \centering
  \includegraphics[width=1.0\linewidth]{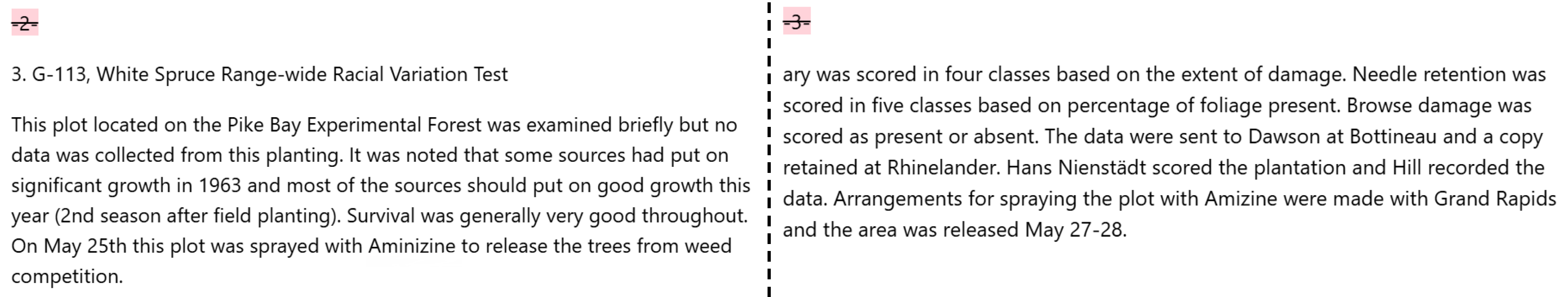}
  \caption{Final state after page number removal. The result shows that the system successfully removed page numbers across the entire document while preserving the rest of the content.}
  \label{fig:section4-2}
\end{figure}

Our system can also perform document-wide modifications, such as removing page numbers or repeated text. For example, the instruction \textit{``Do not put the page number text like `-2-' in the output''} automatically removed similar artifacts across multiple sections while preserving the remaining content. The original documents and extractions are shown in \autoref{fig:section4-1}, and the refined result is shown in \autoref{fig:section4-2}.

\begin{figure}[htbp]
  \centering
  \includegraphics[width=1.0\linewidth]{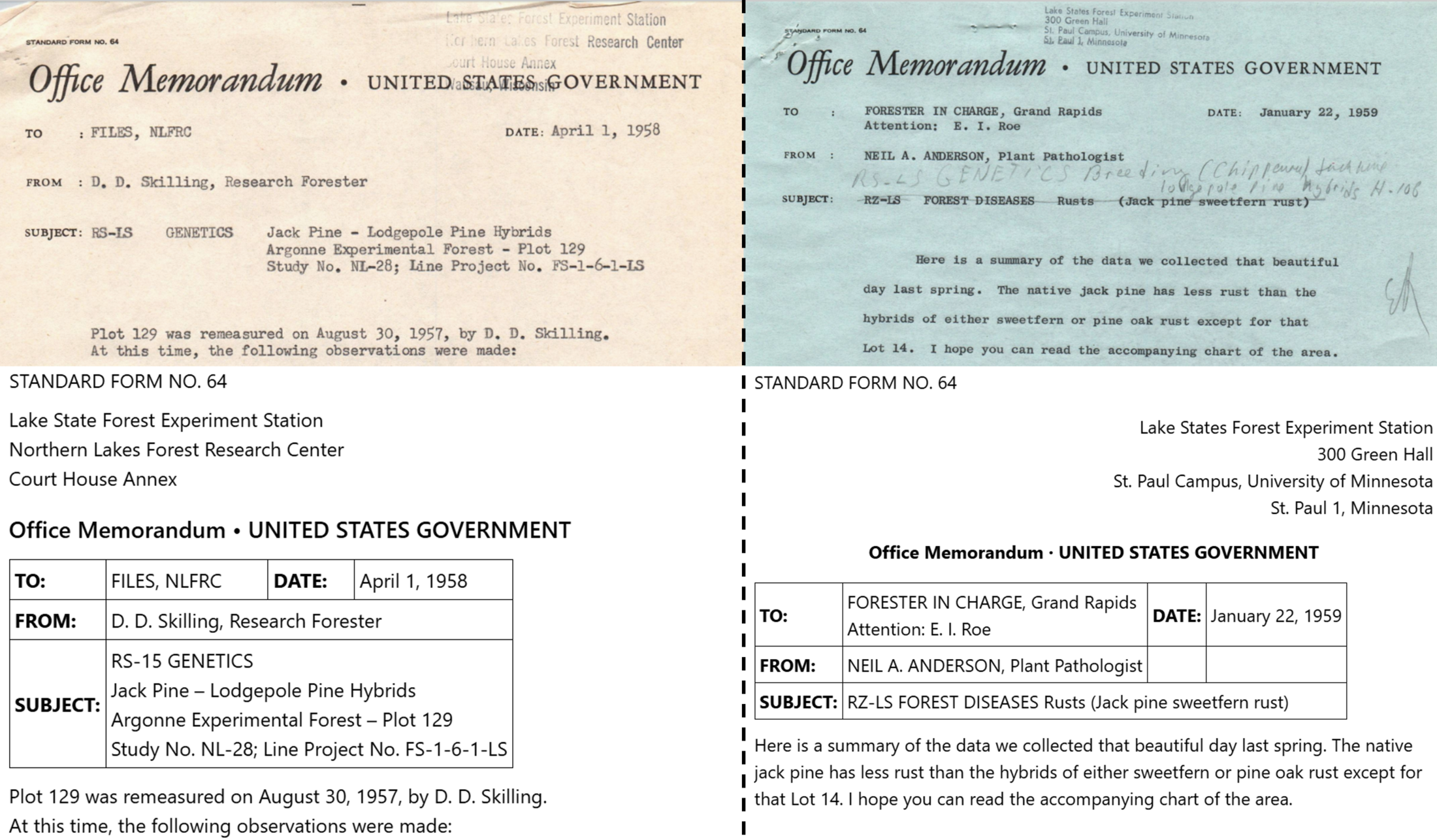}
  \caption{Initial state of repeated text removal. The uploaded documents (top) contain institutional headers, such as ``STANDARD FORM NO. 64,'' repeated across many pages. In the extracted version (bottom), these repeated elements are also present.}
  \label{fig:section4-3}
\end{figure}

\begin{figure}[htbp]
  \centering
  \includegraphics[width=1.0\linewidth]{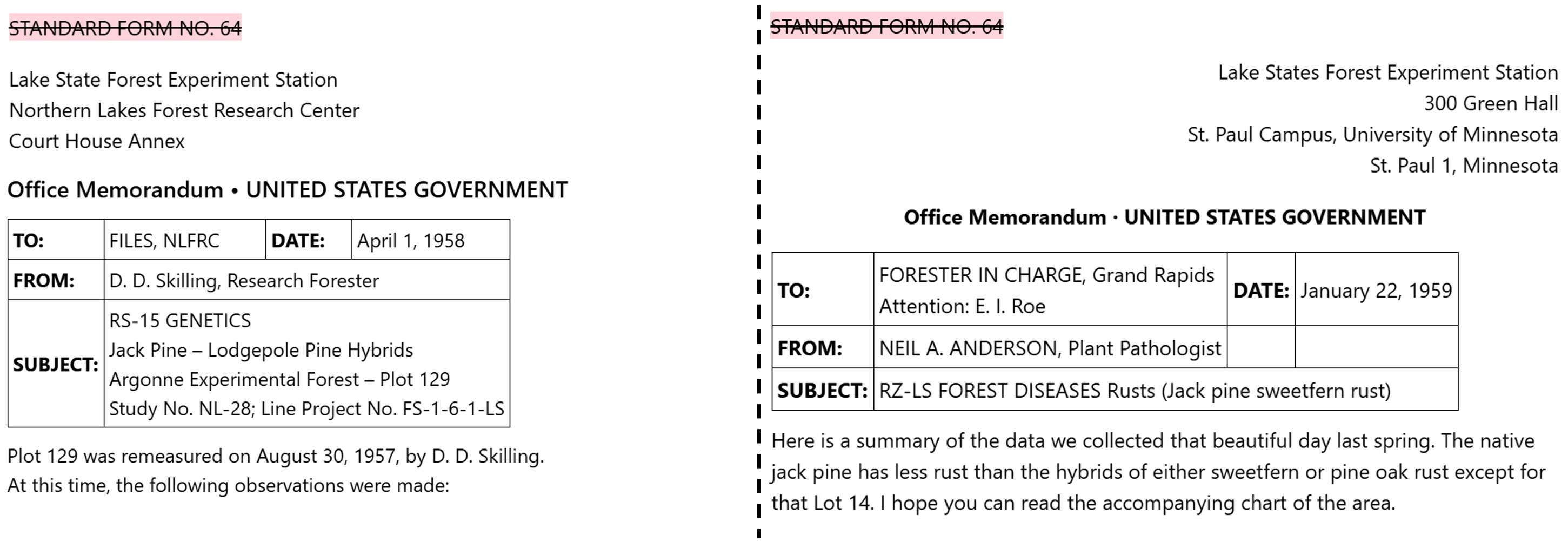}
  \caption{Result of repeated text removal. The system successfully eliminated redundant elements such as ``STANDARD FORM NO. 64'' that appeared across the document. This refinement reduces clutter and improves readability by retaining only meaningful content.}
  \label{fig:section4-4}
\end{figure}

In addition, the system can identify and remove repeated phrases across an entire document. For example, the documents in \autoref{fig:section4-3} contain repeated institutional headers across several pages. The instruction \textit{``remove `STANDARD FORM NO. 64' across the whole document''} removed these repeated elements throughout the document. The result is shown in \autoref{fig:section4-4}.

\subsection{Example-Guided Structural Correction}

\begin{figure}[htbp]
  \centering
  \includegraphics[width=0.4\linewidth]{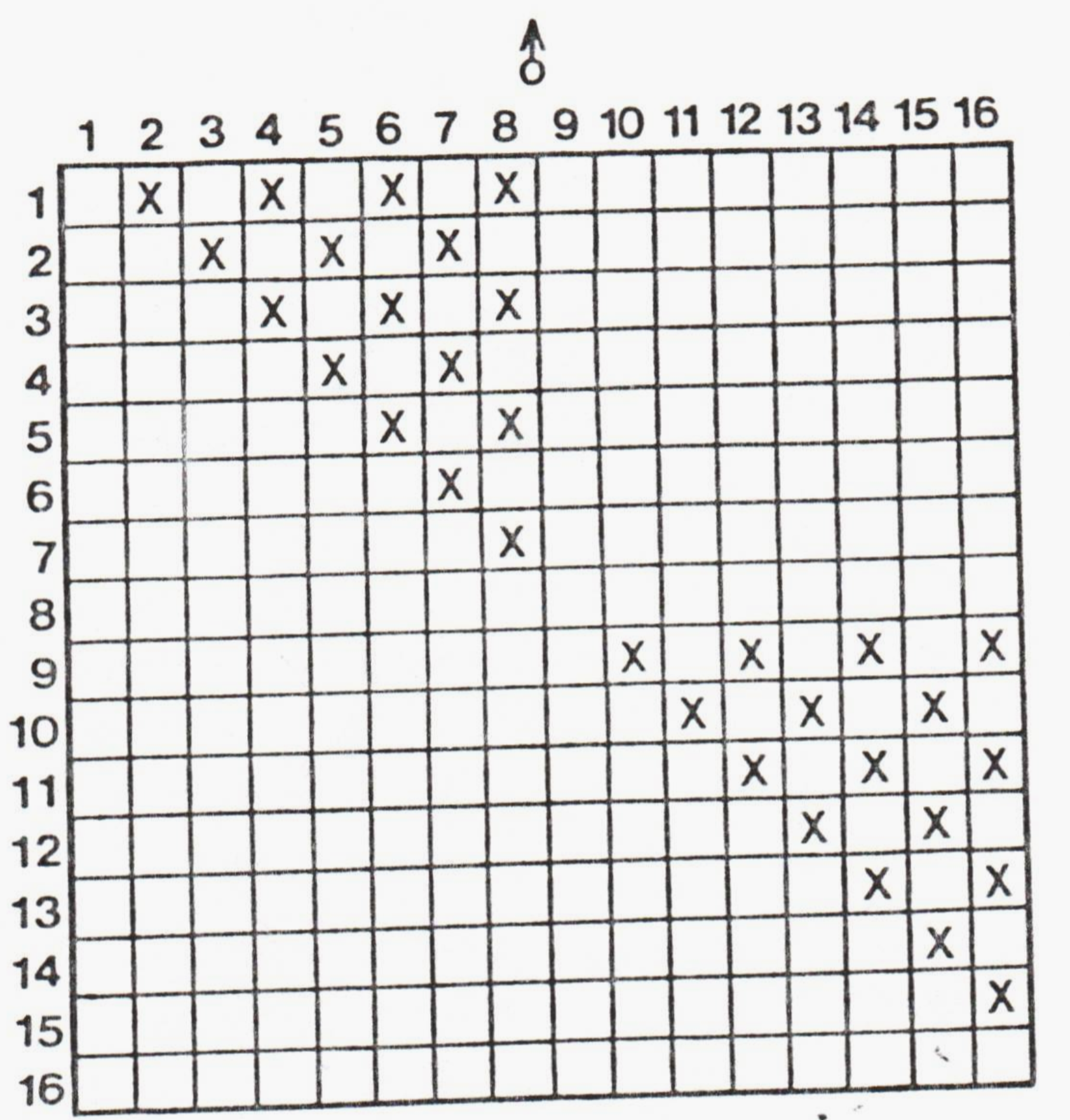}
  \caption{Uploaded document illustrating a case of example-guided structural correction. The image shows a $16 \times 16$ square table with several cells containing the letter \textit{``X''}.}
  \label{fig:section5-1}
\end{figure}

\begin{figure}[htbp]
  \centering
  \includegraphics[width=0.6\linewidth]{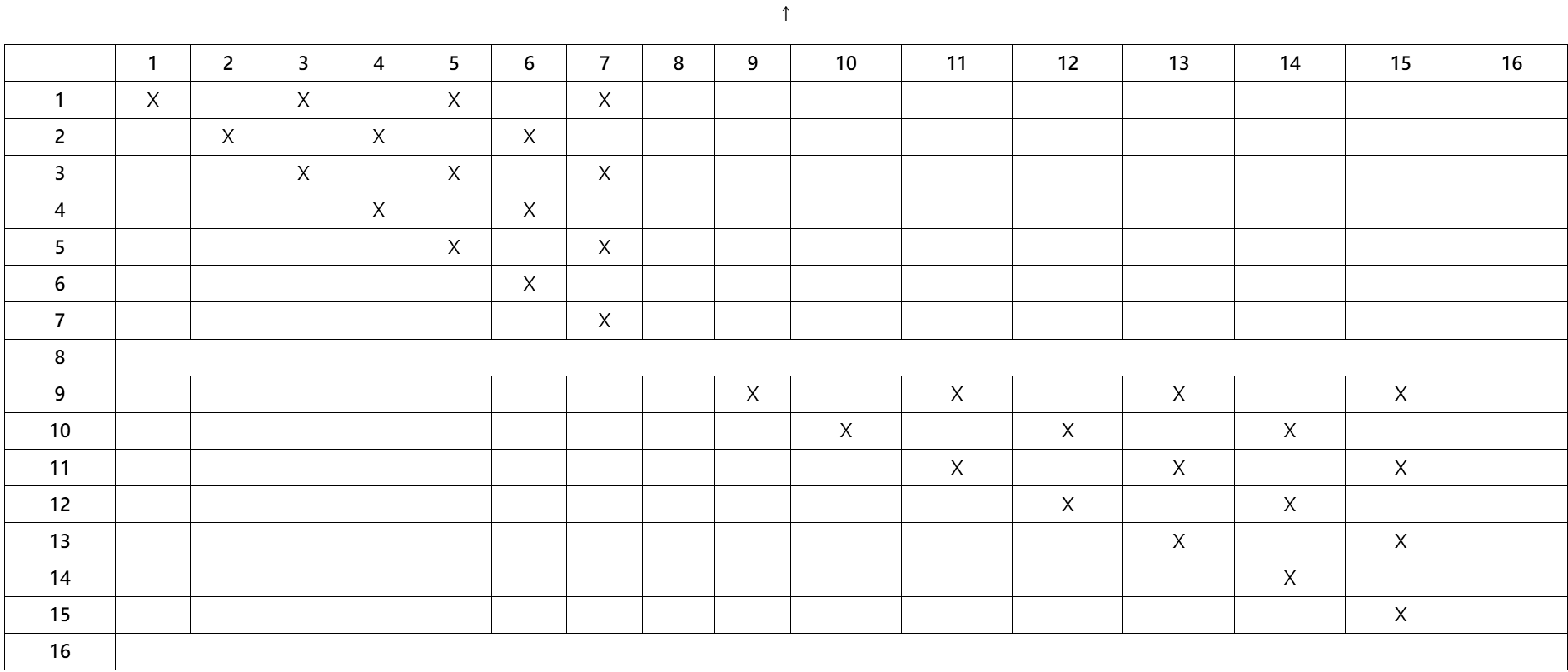}
  \caption{Extracted version of the $16 \times 16$ square table, showing misordered \textit{``X''} symbols across the table.}
  \label{fig:section5-2}
\end{figure}

\begin{figure}[htbp]
  \centering
  \includegraphics[width=0.6\linewidth]{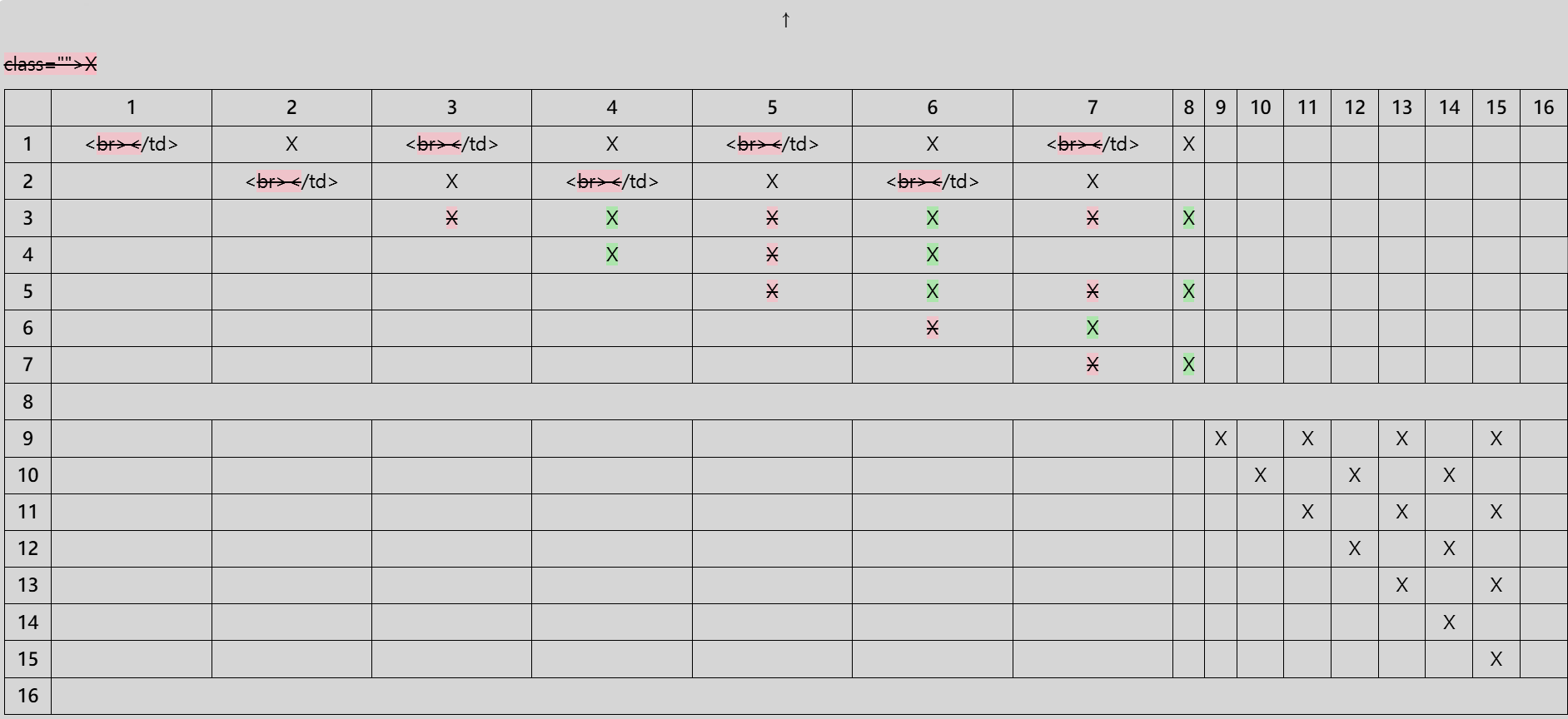}
  \caption{The system learns from the instruction \textit{``you miss type the order of X, I've modified the first 2 rows, look into it and modify the rest X in the correct order''} and follows the modified order provided by the user to correct the remaining \textit{``X''} symbols.}
  \label{fig:section5-3}
\end{figure}

\begin{figure}[htbp]
  \centering
  \includegraphics[width=0.6\linewidth]{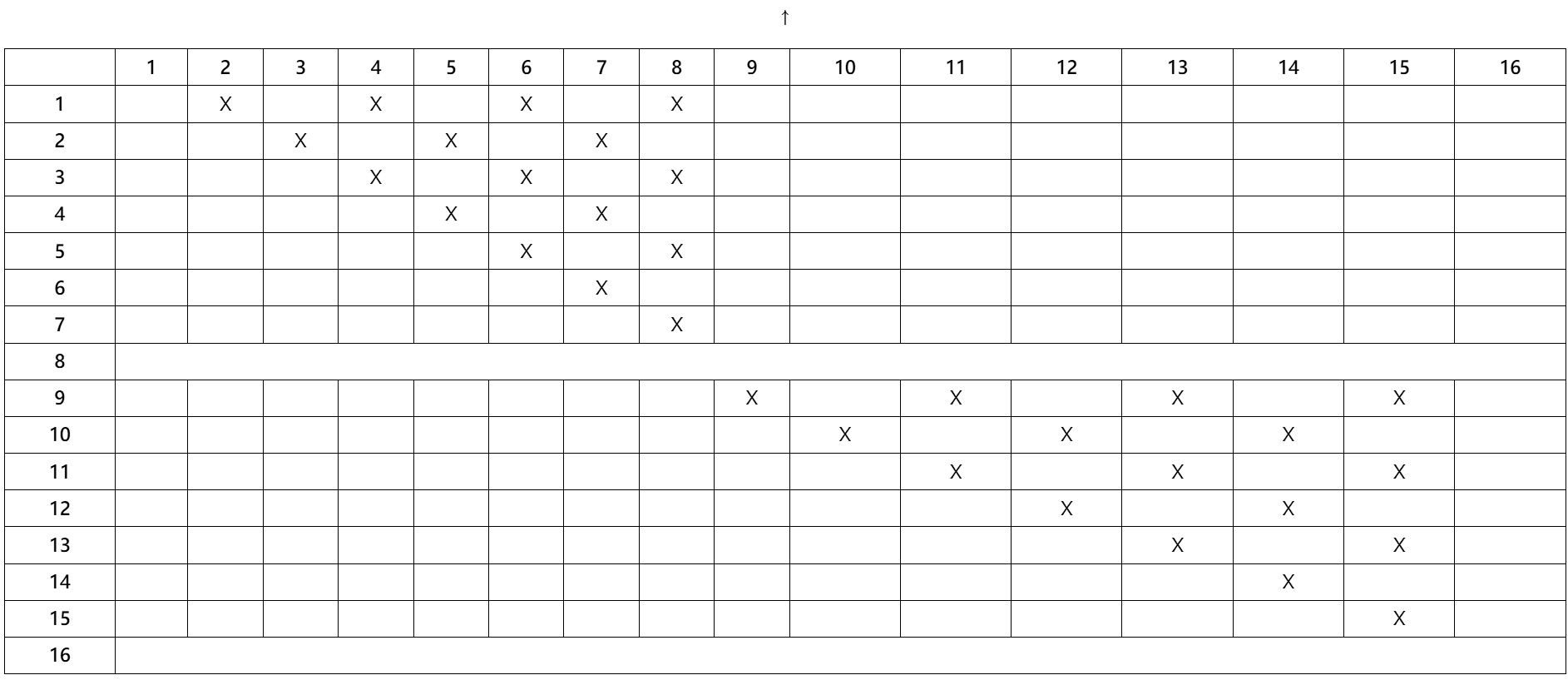}
  \caption{Final state of the example-guided structural correction case. The figure shows the corrected $16 \times 16$ table after the system applied the user's example-based modification and restored the proper order of \textit{``X''} marks across the grid.}
  \label{fig:section5-4}
\end{figure}

In some cases, extracted tables contain structural errors such as misordered symbols. Our system can follow user-provided examples and instructions to reorder symbols across an entire table. In one case, a user observed from the input table shown in \autoref{fig:section5-1} that the extraction had misordered the \textit{``X''} symbols throughout the table, as shown in \autoref{fig:section5-2}. To fix the issue, the user manually edited the first two rows and provided the instruction \textit{``you miss type the order of X, I've modified the first 2 rows, look into it and modify the rest X in the correct order.''} The system then generalized this correction pattern, as illustrated in \autoref{fig:section5-3}, and applied the structural adjustment across the full table, producing the corrected result shown in \autoref{fig:section5-4}. This example demonstrates the system's ability to infer structural rules from user-provided edits.

\subsection{Direct Edits Without Instructions}

\begin{figure}[htbp]
  \centering
  \includegraphics[width=0.5\linewidth]{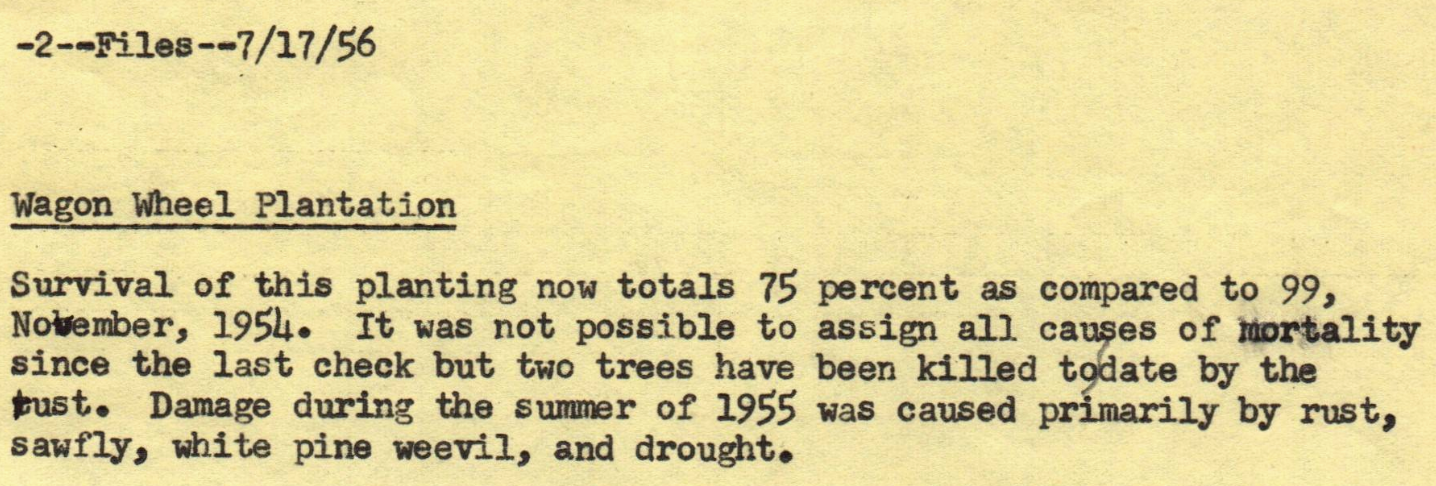}
  \caption{Uploaded document illustrating a case of direct edits without instructions. The image contains the misspelled word \textit{``todate''} within the content.}
  \label{fig:section6-1}
\end{figure}

\begin{figure}[htbp]
  \centering
  \includegraphics[width=1.0\linewidth]{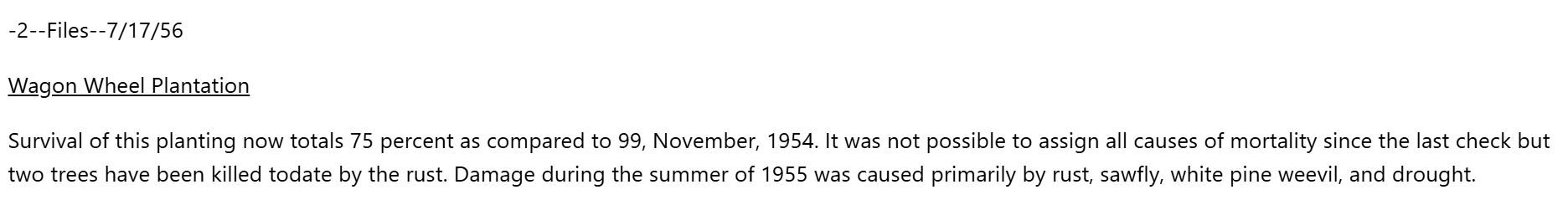}
  \caption{Extracted version of the document, where the misspelled word \textit{``todate''} is visible in the content.}
  \label{fig:section6-2}
\end{figure}

\begin{figure}[htbp]
  \centering
  \includegraphics[width=1.0\linewidth]{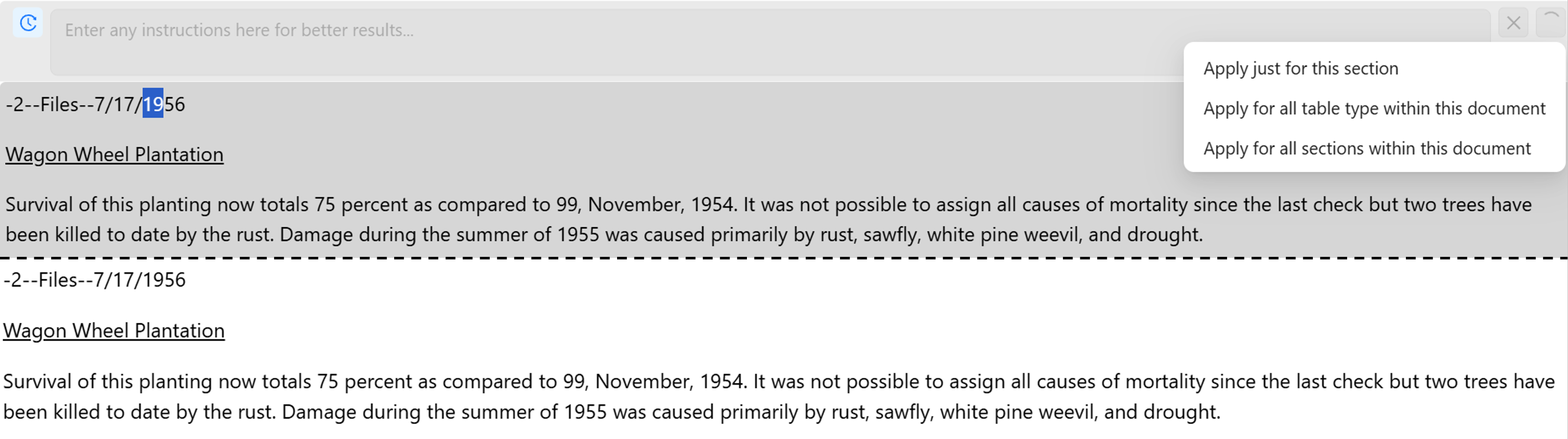}
  \caption{Final state of the direct edit case. The figure shows how a user directly edits the extracted text, including changing the year to \textit{``1956''} and correcting the mistyped word \textit{``todate''} to \textit{``to date.''} The upper image shows the editing interface, and the lower image shows the final result.}
  \label{fig:section6-3}
\end{figure}

Our system also supports direct modification of extracted content without requiring users to enter prompts. In one case, the user directly edited a misspelled document shown in \autoref{fig:section6-1}, with the extracted version shown in \autoref{fig:section6-2}, by changing the date from \textit{``7/17/56''} to \textit{``7/17/1956''} and adding a space in the misspelled word \textit{``todate.''} The final result is shown in \autoref{fig:section6-3}, where the user successfully corrected both the date and the misspelled word through direct editing alone.

\end{document}